\documentclass[12pt]{article}
\pdfoutput=1
\usepackage{graphicx}
\usepackage{mathrsfs}
\usepackage{amssymb}
\usepackage{amsmath}
\usepackage{graphicx}
\usepackage{amsmath,amsfonts}
\usepackage{algorithm}
\usepackage{algorithmicx}
\usepackage{algpseudocode}
\usepackage{amscd}
\usepackage[mathscr]{eucal}
\usepackage{lineno}
\usepackage{extarrows}
\usepackage{url}

\usepackage{listings}
\usepackage{tikz}

\newcommand{\pkg}[1]{\textbf{#1}}
\newcommand{\proglang}[1]{\texttt{#1}}
\newcommand{\code}[1]{\texttt{\textit{#1}}}

\usepackage{tikz}

\numberwithin{thm}{section}

\newcommand{\be}{\begin{equation}} \newcommand{\ee}{\end{equation}}
\newcommand{\bd}{\begin{displaymath}} \newcommand{\ed}{\end{displaymath}}
\newcommand{\ba}{\begin{align}} \newcommand{\ea}{\end{align}}
\newcommand{\baa}{\begin{align*}} \newcommand{\eaa}{\end{align*}}
\newcommand{\ben}{\begin{enumerate}} \newcommand{\een}{\end{enumerate}}
\newcommand{\bi}{\begin{itemize}} \newcommand{\ei}{\end{itemize}}

\newcommand{\ud}{\mathrm{d}}

\algnewcommand\And{\textbf{and}}

\begin{document}


\title{Modelling trait dependent speciation with Approximate Bayesian Computation}
\author{Krzysztof Bartoszek and Pietro Li\`o} 

\maketitle

\begin{abstract}
Phylogeny is the field of modelling the temporal discrete dynamics of
speciation. Complex models can nowadays be studied using the Approximate
Bayesian Computation approach which avoids likelihood calculations. The field's
progression is hampered by the lack of robust software to estimate the
numerous parameters of the speciation process.
In this work we present an \proglang{R} package, \pkg{pcmabc}, based
on Approximate Bayesian Computations, that implements three novel phylogenetic
algorithms for trait--dependent speciation modelling.
Our phylogenetic
comparative methodology takes into account both the simulated traits and
phylogeny, attempting to estimate the parameters of the processes
generating the phenotype and the trait. The user is not restricted to a
predefined set of models and can specify a variety of evolutionary and
branching models. We illustrate the software with a simulation--reestimation
study focused around the branching Ornstein--Uhlenbeck process, where the
branching rate depends non--linearly on the value of the driving
Ornstein--Uhlenbeck process. Included in this work is a tutorial on how to
use the software.
\end{abstract}


\section{Introduction}

The relationship between genotypes and phenotypes originates from a very complex spatial and temporal, non--linear 
dynamical system. Due to the quick and microscopic dynamics, parts of the developmental process are usually hidden 
from direct observation. Data on adult animals are rarely collected in a continuous manner. The characteristics of a 
phenotype depend on both genetic and environmental factors so that genetically identical individuals could have a
different phenotype expressively due to the subtle dependency on environmental (exposomes) factors.

Phylogenetics occupies an extremely important position in the Modern Synthesis and it is central to 
most areas of biology. It 
bridges population genetics \cite{KStrOPyb2001}, genomics and cancer research \cite{ASotetal2013}.
Furthermore, there is hope that it may assist in discovering relationships between
complex diseases \cite{HXiaKBarPLio2018}.

In the last decade, the area of phylogenetics has been tremendously exposed to novel statistical and computational models 
previously adopted only in theoretical studies. Bayesian methodologies are now at the core of phylogenetic comparative 
methods (PCM---the study of phenotypic data on the between--species level, the trait measurements 
are dependent through the species' common evolutionary history) and are used to evaluate macroevolutionary hypotheses of phenotypic evolution under distinct evolutionary 
processes in a phylogenetic context. 

In particular, Approximate Bayesian Computation (ABC) is a powerful methodology to estimate the posterior distributions 
of model parameters without evaluating the (usually computationally very costly) likelihood function
\cite{PDigRGra1984,DRub1984}. That property has 
widened the domains of application of Bayesian models 
(see e.g. \cite{JLinMGutRDutSKasJCor2017} for recent developments of ABC approaches for evolutionary biology)
and has offered interesting challenges in parameters estimation tasks.
This leads to an interesting consideration: biology, in particular DNA sequence analysis, has been central for the 
development of the ABC methodology \cite{STavetal1997}, in return biology represents a rich application domain for 
Bayesian analysis which could probably inspire further theoretical developments. 

Another key factor for these new Bayesian theoretical frameworks has been the statistical computing language \proglang{R} \cite{R}
that is central to a community of scientists who have developed a wide range of tools and functions for phylogenetic 
comparative analyses. The development of tools (see for instance
\url{https://cran.r-project.org/web/views/Phylogenetics.html}) has grown in parallel with the perception of how 
the use of newly developed tools would facilitate the understanding of statistical and biological concepts, 
produce successful instances of phylogenetic inference and bridge the gap between mathematicians and biologists. 
Therefore, the large existing variety of tools reflects and accommodates the current rich interdisciplinary and 
multidisciplinary field of phylogenetic studies. To a beginner, phylogenetics as a field, 
represents the intersection of interesting questions, great data and powerful tools. 

With the above considerations in mind, here we would like to address 
trait dependent speciation models by means of phylogeny--based evolutionary dynamics. The vastity and complexity of the 
research theme is attacked by means of a general, although focused, tool based on powerful ABC approach. The beginner is 
accurately guided by a tutorial, a description of the package and by a set of modelling questions and case studies.

The paper has the following structure. In the next Section, \ref{secpcmabc}, we introduce the \pkg{pcmabc} \proglang{R} package and the 
algorithms for simulation of traits and trees. 
In particular in Section \ref{secpcmabc} we describe three novel phylogenetic algorithms required
by the ABC inference procedure.
Then, in Section \ref{secSimul} we describe a simulation study to evaluate 
whether the ABC 
inference package can capture any signal on the trait dependent speciation process, based only on the contemporary 
sample and phylogeny. The tutorial in Section \ref{secTutorial} is self contained although it uses the 
same Ornstein--Uhlenbeck (OU) process as in Section \ref{secSimul}. 
We end the paper with Section \ref{secConc} which summarizes the possibilities
of the software, its limitations and possible directions of future development.
Although we leave to the reader the choice of the order, we are delighted to report that the tool is not only serving 
the scope of studying by simulations the theory on evolution of traits or rapid prototyping the implementation of new 
hypotheses. It embeds a large generality towards a class of problems at the core of today's  theoretical advancements in 
phylogenetics. Therefore, we aim in the future at collecting and presenting in a website the parameters and 
the biological results obtained using this methodology and software.  

To facilitate reading, we have adopted the following fonts for the computer code. Programming language
names are written in  typewriter font (e.g. \proglang{R}), \proglang{R} package names
are written in bold (e.g. \pkg{pcmabc}) and inline code is written in italicized typewriter
font (e.g. \code{x<-1}).

\begin{figure}[!ht]
\centering
\includegraphics[width=9cm]{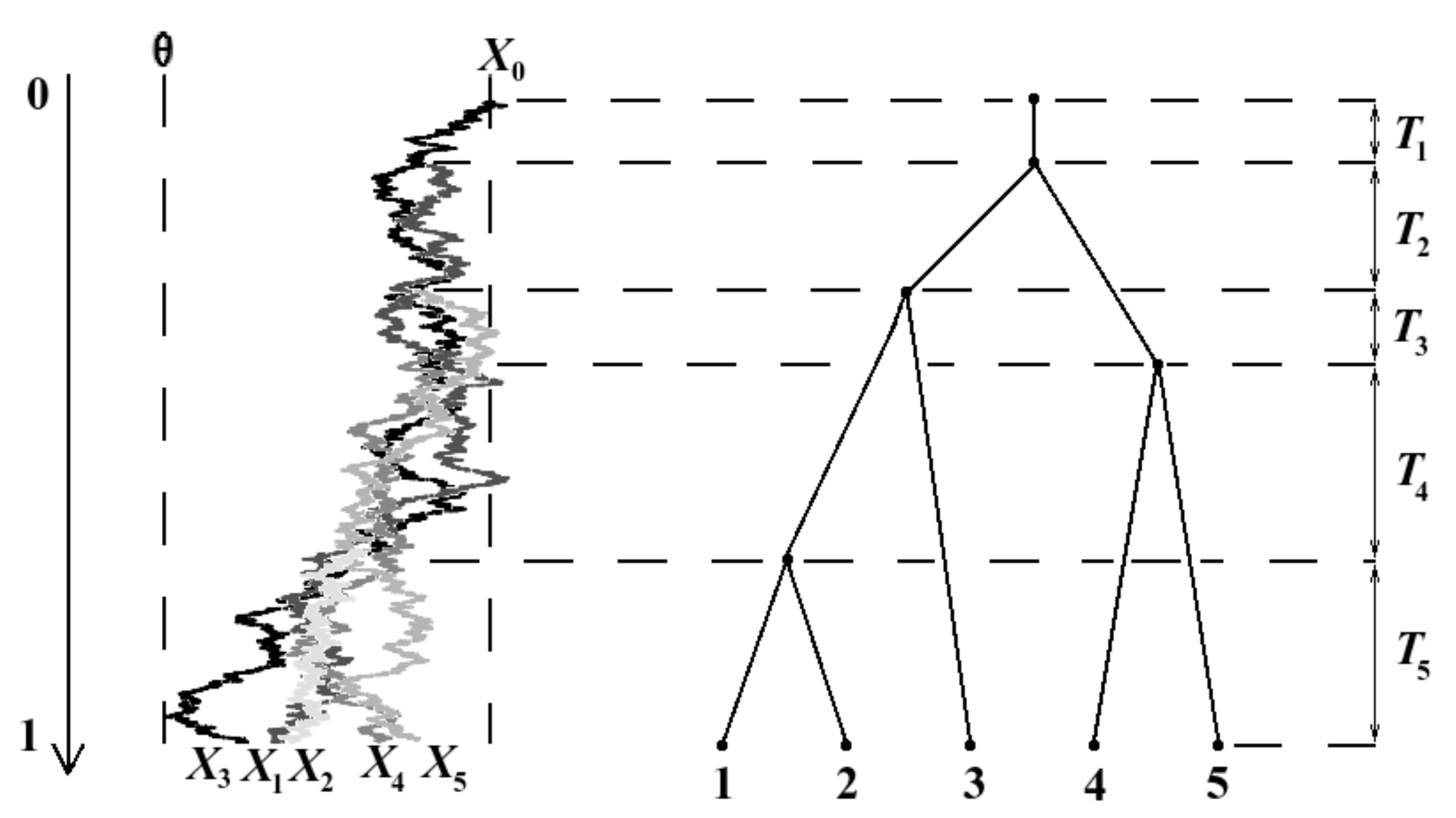}
\caption{
On the left: a branching OU process simulated on a realization of a tree with $n=5$ tips
using the 
\pkg{TreeSim} \cite{TreeSim1,TreeSim2} and \pkg{mvSLOUCH} \cite{KBaretal2012} \proglang{R}  packages. 
Parameters used are $\alpha=1$, $\sigma=1$,  $X_0-\theta=2$, after the tree height was scaled to height $1$.
On the right: the species tree disregarding the trait values supplied with the notation for the inter--speciation times. 
Reprinted by permission of the Applied Probability Trust. First published in Journal of Applied Probability $52(4)$. Copyright \textcopyright Applied Probability Trust 2015
}
\label{tr}
\end{figure}

\begin{figure}[!ht]
\begin{center}
{
\includegraphics[width=0.243\textwidth]{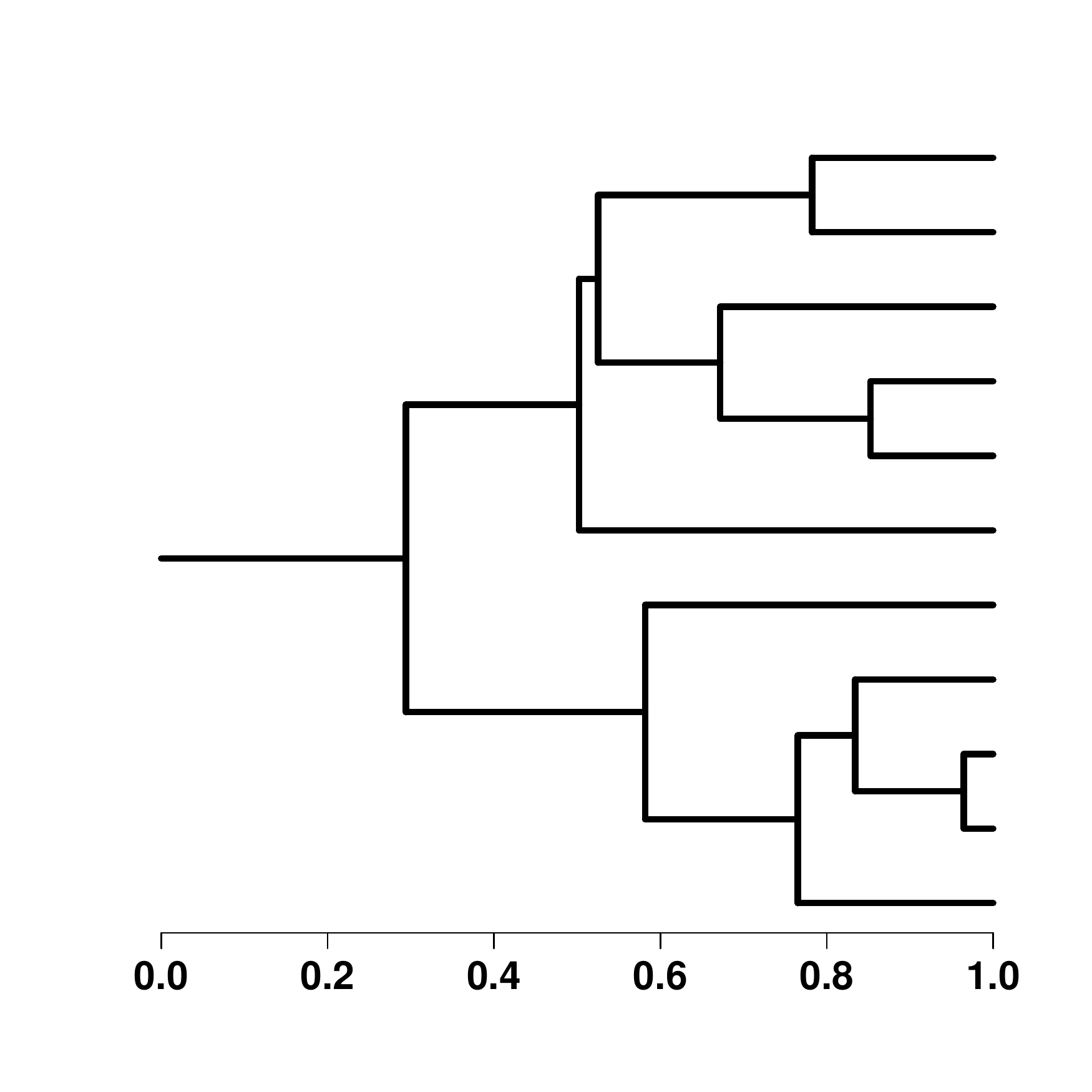}
\includegraphics[width=0.243\textwidth]{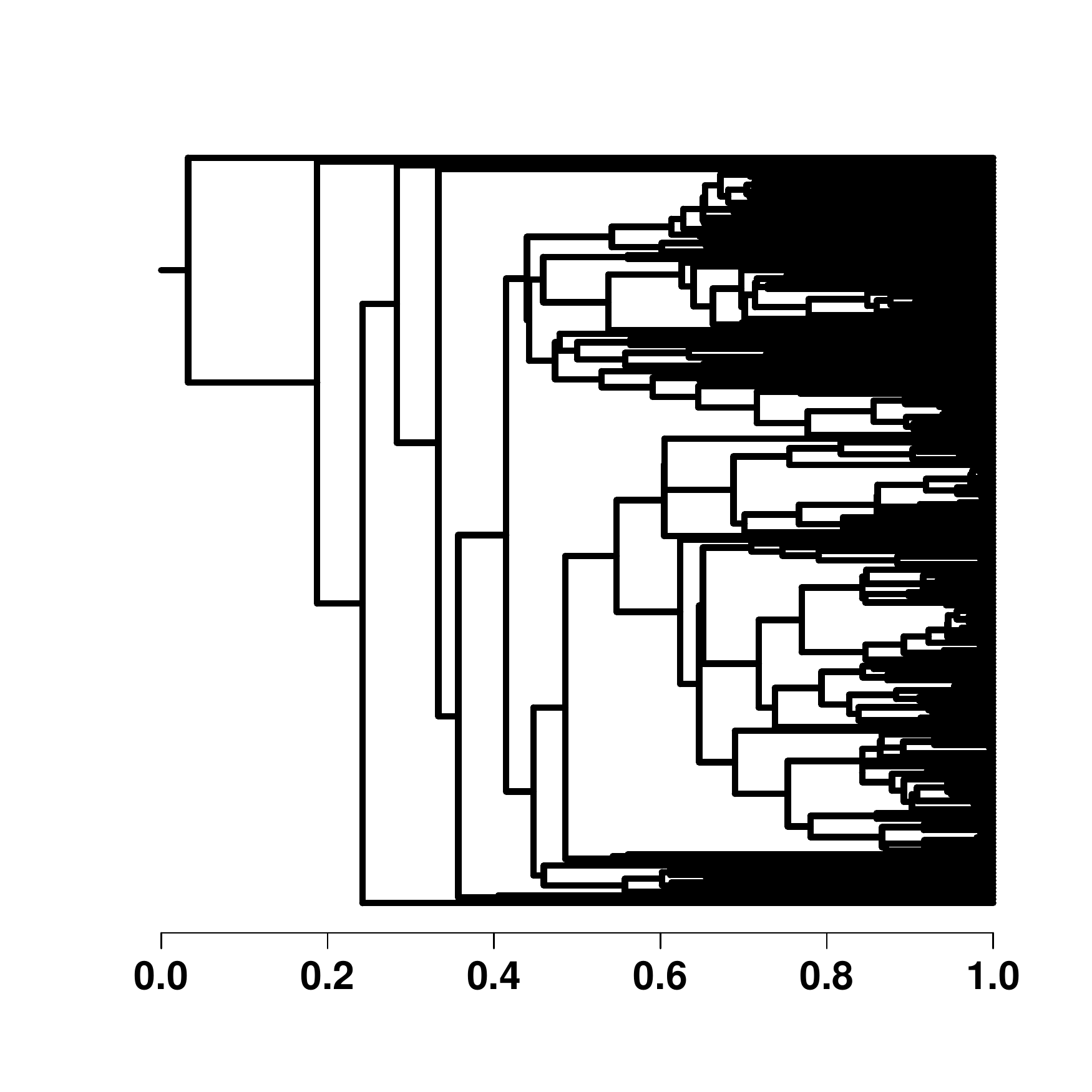}
\includegraphics[width=0.243\textwidth]{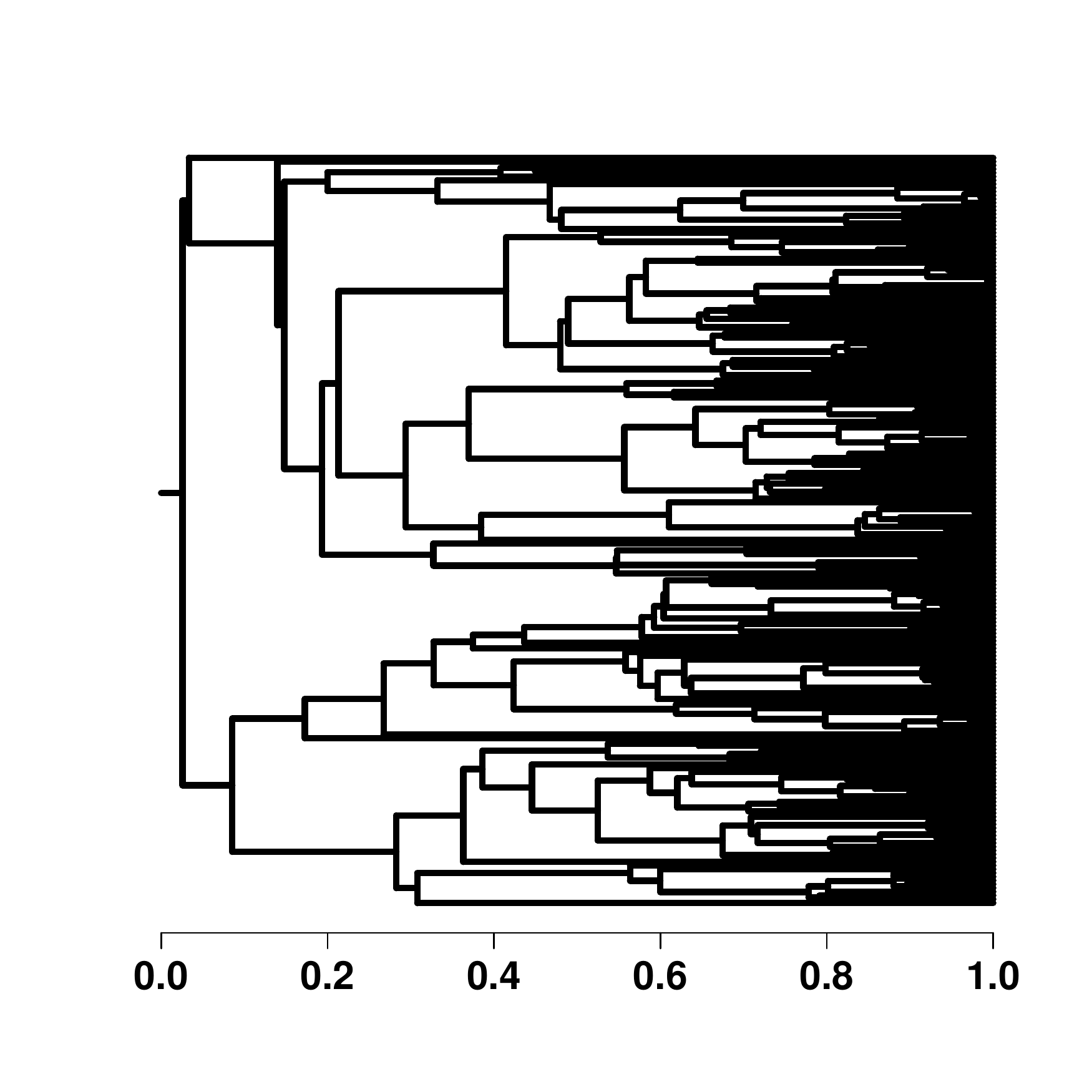}  
\includegraphics[width=0.243\textwidth]{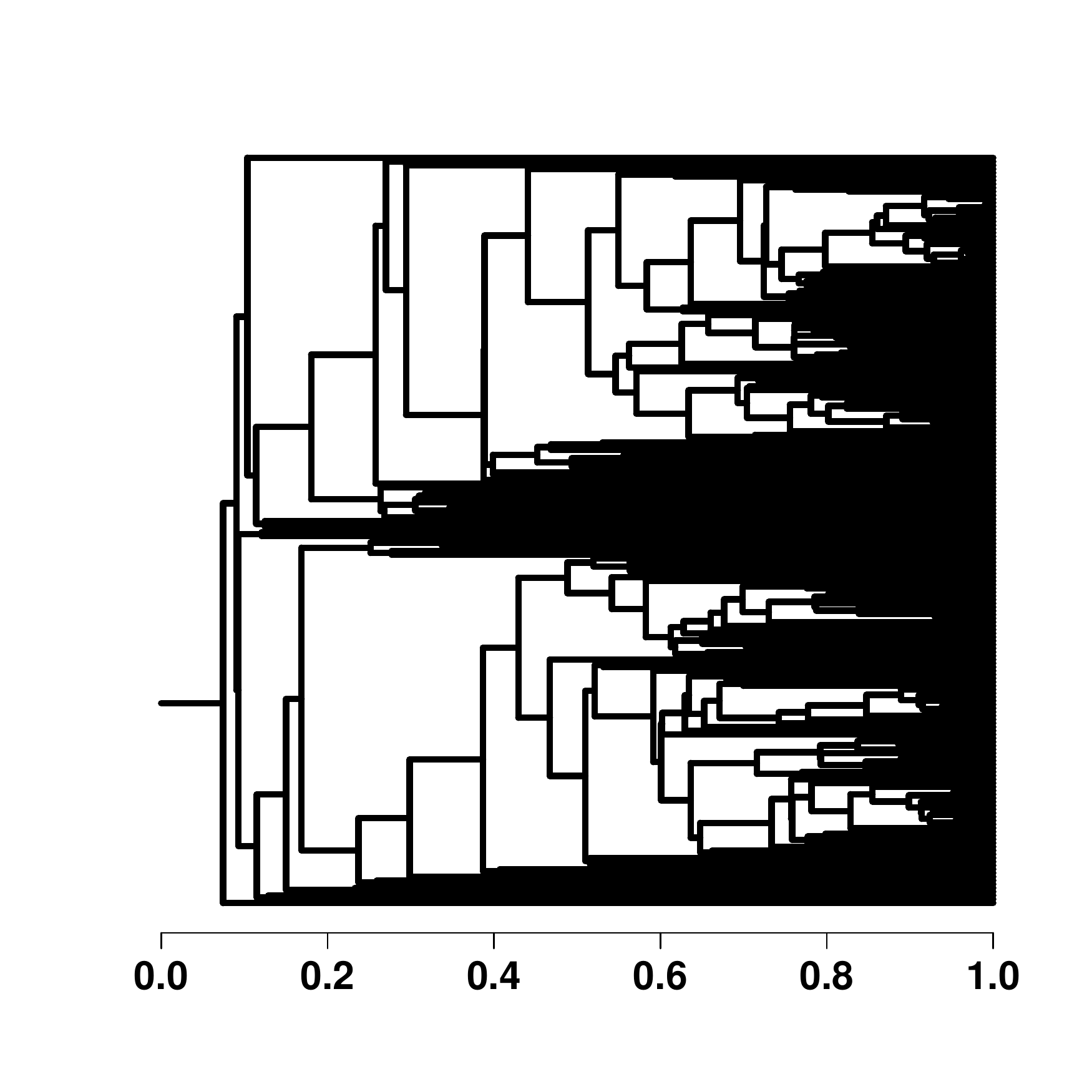} \\
\includegraphics[width=0.242\textwidth,angle=90]{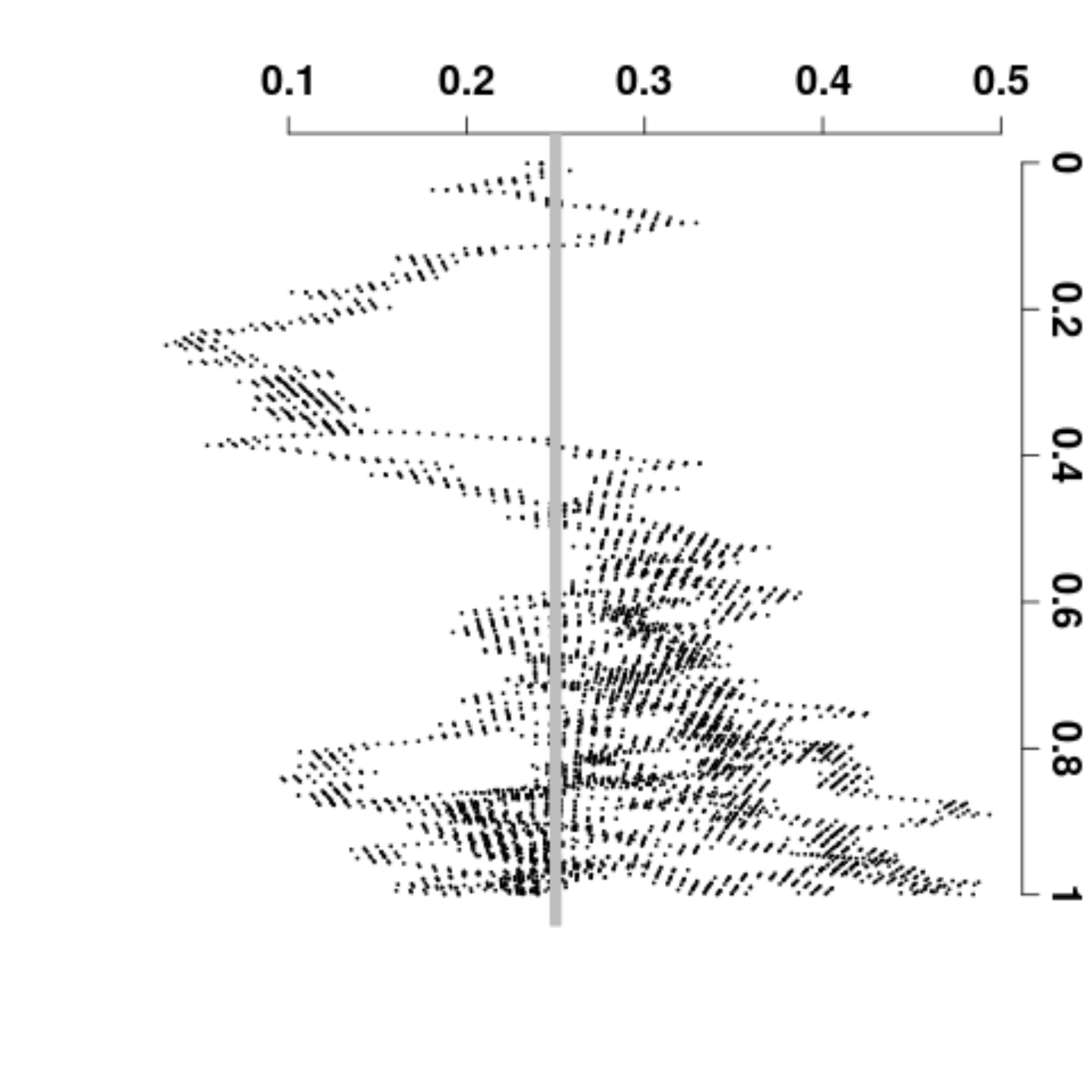}
\includegraphics[width=0.242\textwidth,angle=90]{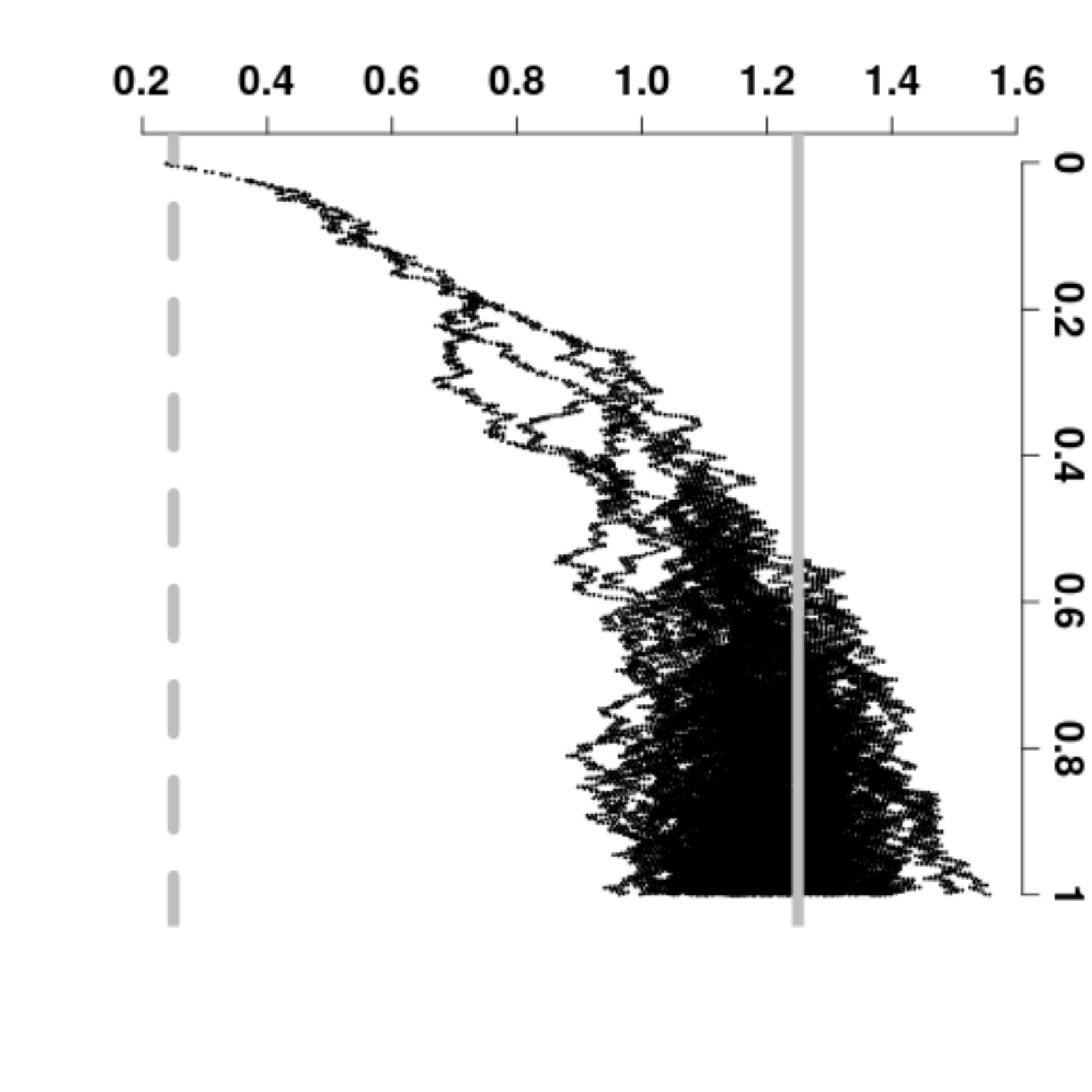} 
\includegraphics[width=0.242\textwidth,angle=90]{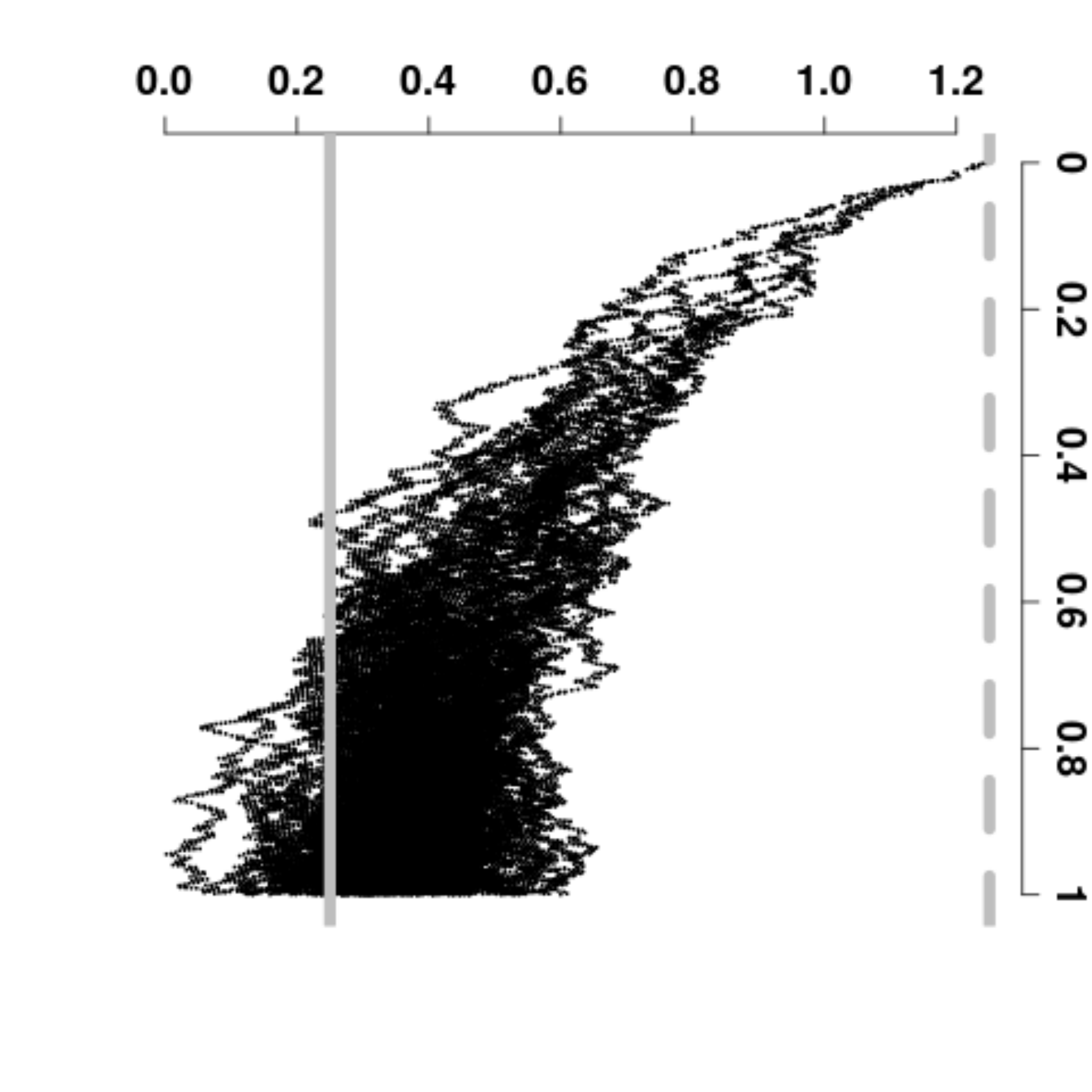}
\includegraphics[width=0.242\textwidth,angle=90]{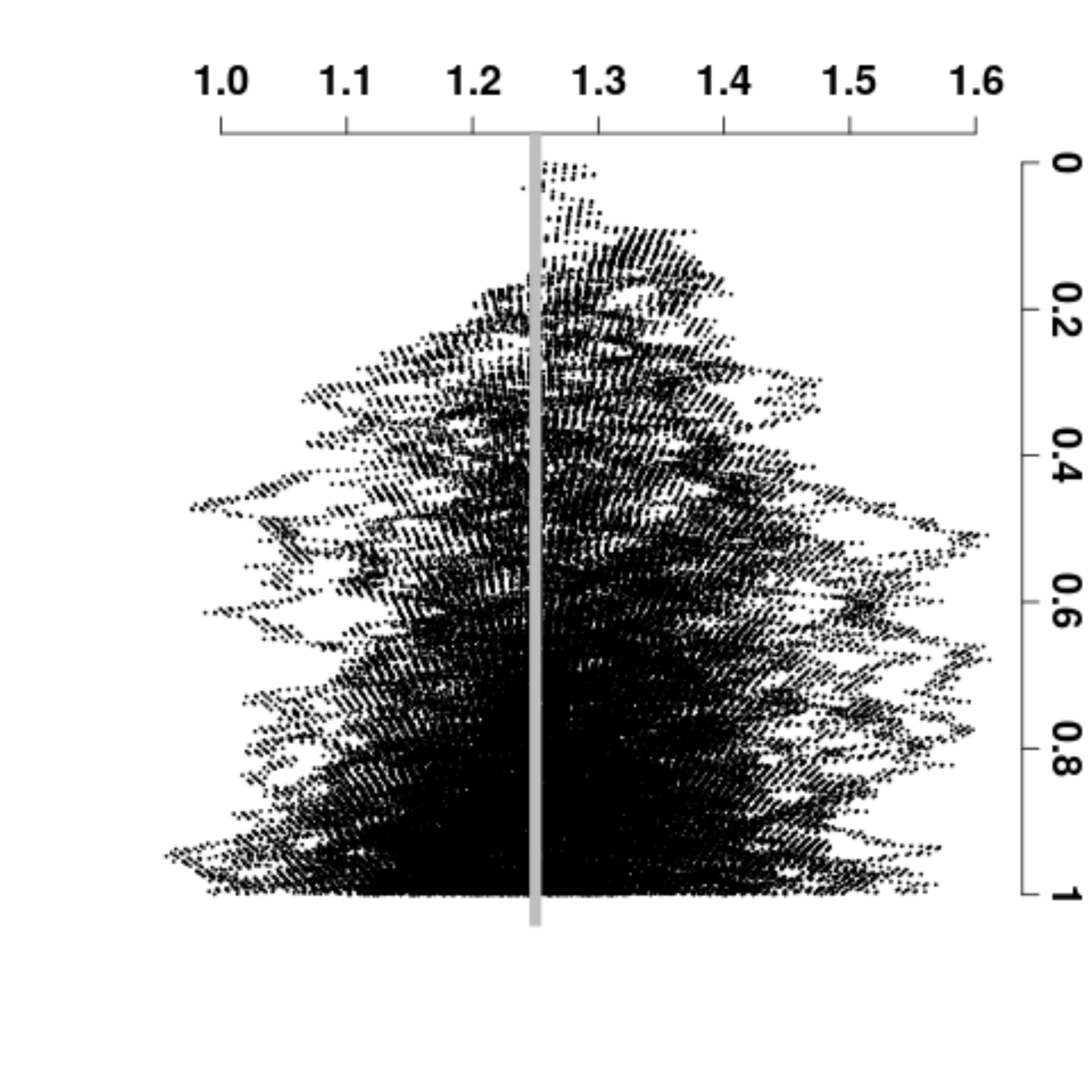}  
}
\end{center}
\caption{Examples of phylogenies (top row) and trait trajectories (second row)
generated by trait dependent speciation models.
The phenotype follows an OU process,
$\ud X_{t} = -3(X_{t} - \psi) \ud t+ 0.25 \ud B_{t}$ and the birth rate 
is $10\cdot \vert \sin(X_{t})\vert$. The choice of the birth rate is for illustrative 
and not biological purposes, i.e. we want speciation dynamics to visibly follow
trait dynamics. It is important for the reader to remember that the phylogenies and traits
are simulated jointly and not that the trait is simulated conditional on the tree, as is 
the case with most PCM simulations.
Columns from left to right: $(X_{0}=0.25,\psi=0.25)$,
$(X_{0}=0.25,\psi=1.25)$, $(X_{0}=1.25,\psi=0.25)$,
$(X_{0}=1.25,\psi=1.25)$.
The simulations were done by the presented here \pkg{pcmabc} package.
The trajectories are plotted using the function \code{pcmabc::draw\_phylproc()}, but without the axes,
these have to be manually added by the user. 
Note that the vertical axis of the trait trajectories are \textbf{NOT} to scale. The height of all trees is $1$.
The stationary mean (solid) and ancestral (dashed) values are marked with gray lines. If there is 
only one line it is because they are equal.
}
\label{figExampleTDepSpec_1}
\end{figure}

\begin{figure}[!ht]
\begin{center}
{
\includegraphics[width=0.242\textwidth,angle=0]{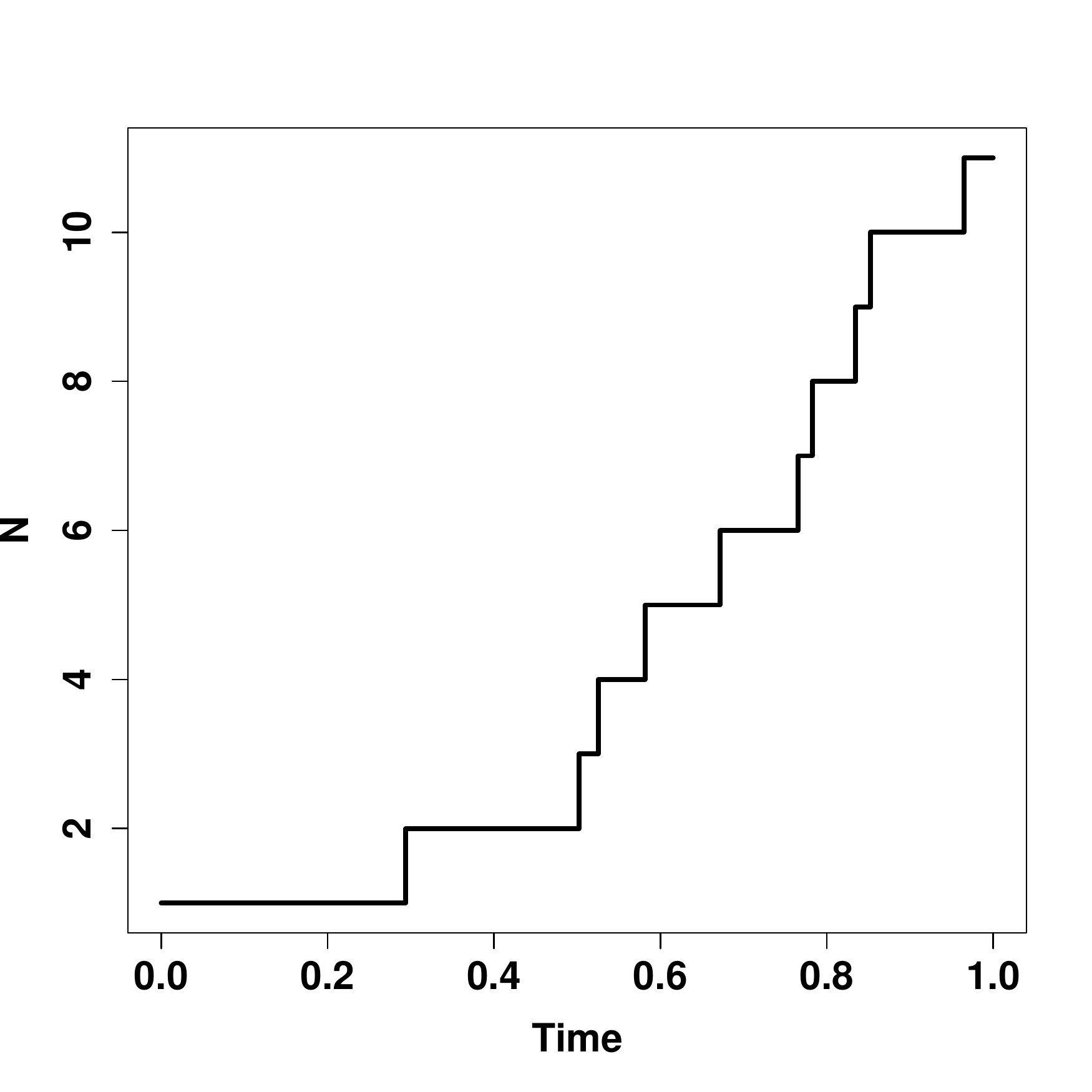}
\includegraphics[width=0.242\textwidth,angle=0]{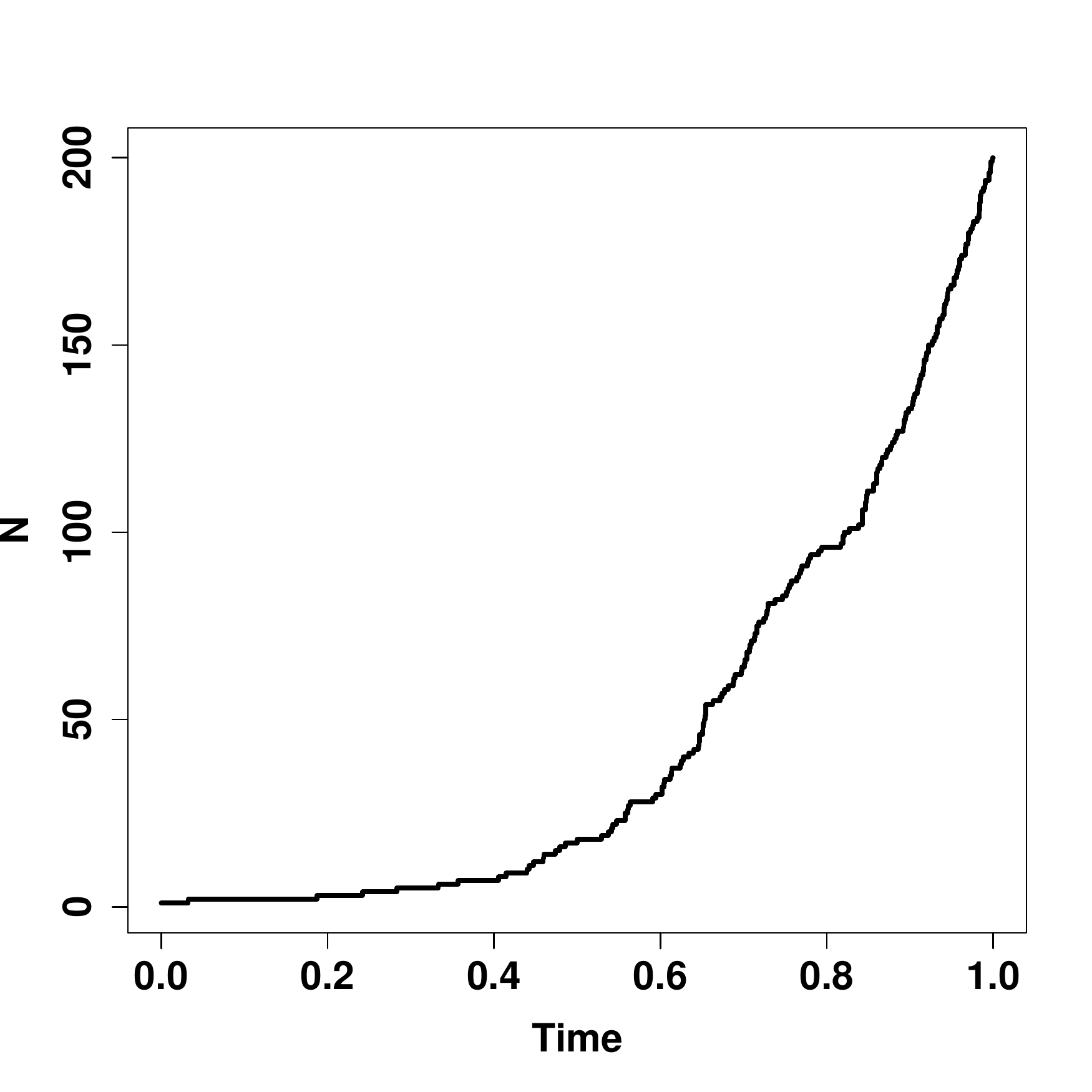} 
\includegraphics[width=0.242\textwidth,angle=0]{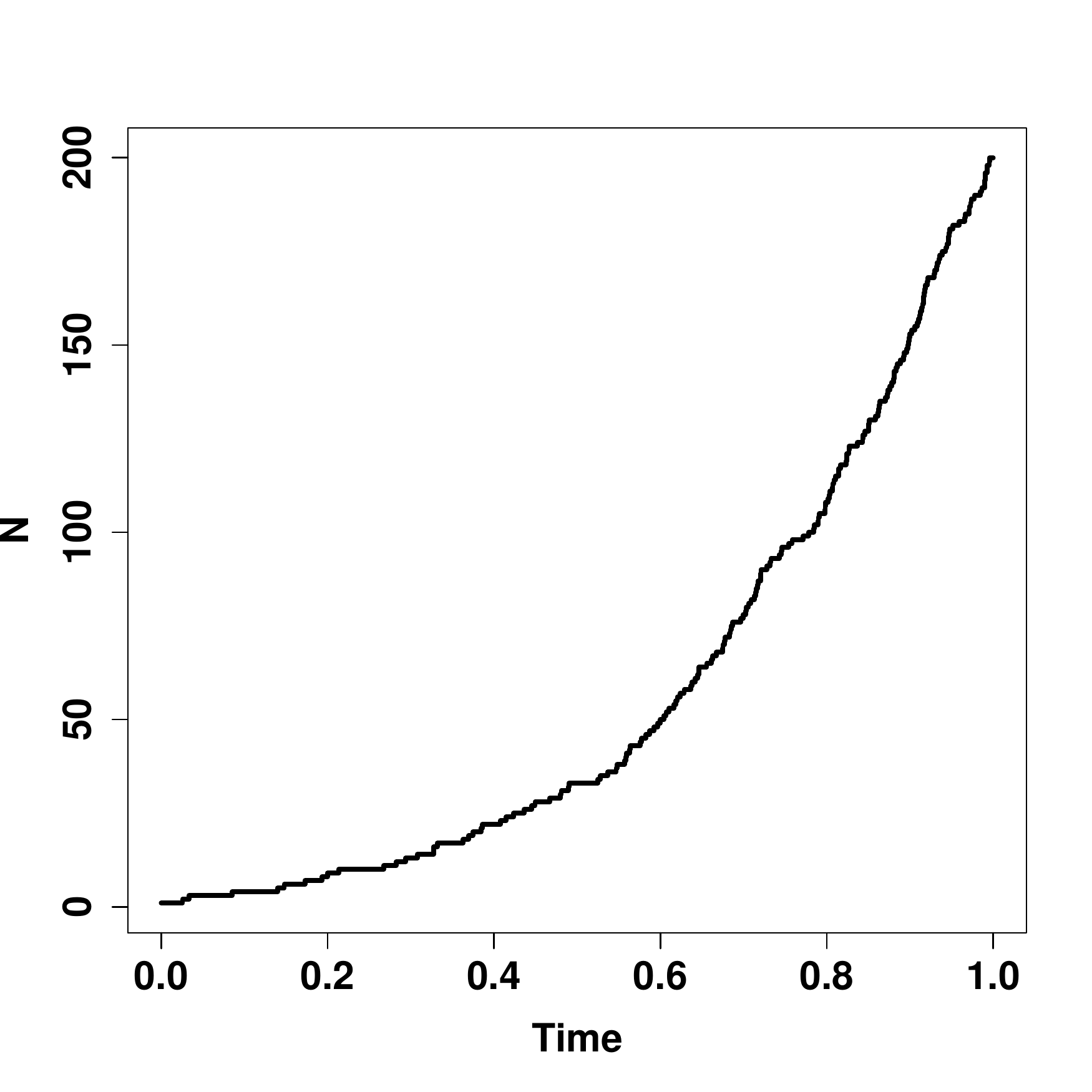}
\includegraphics[width=0.242\textwidth,angle=0]{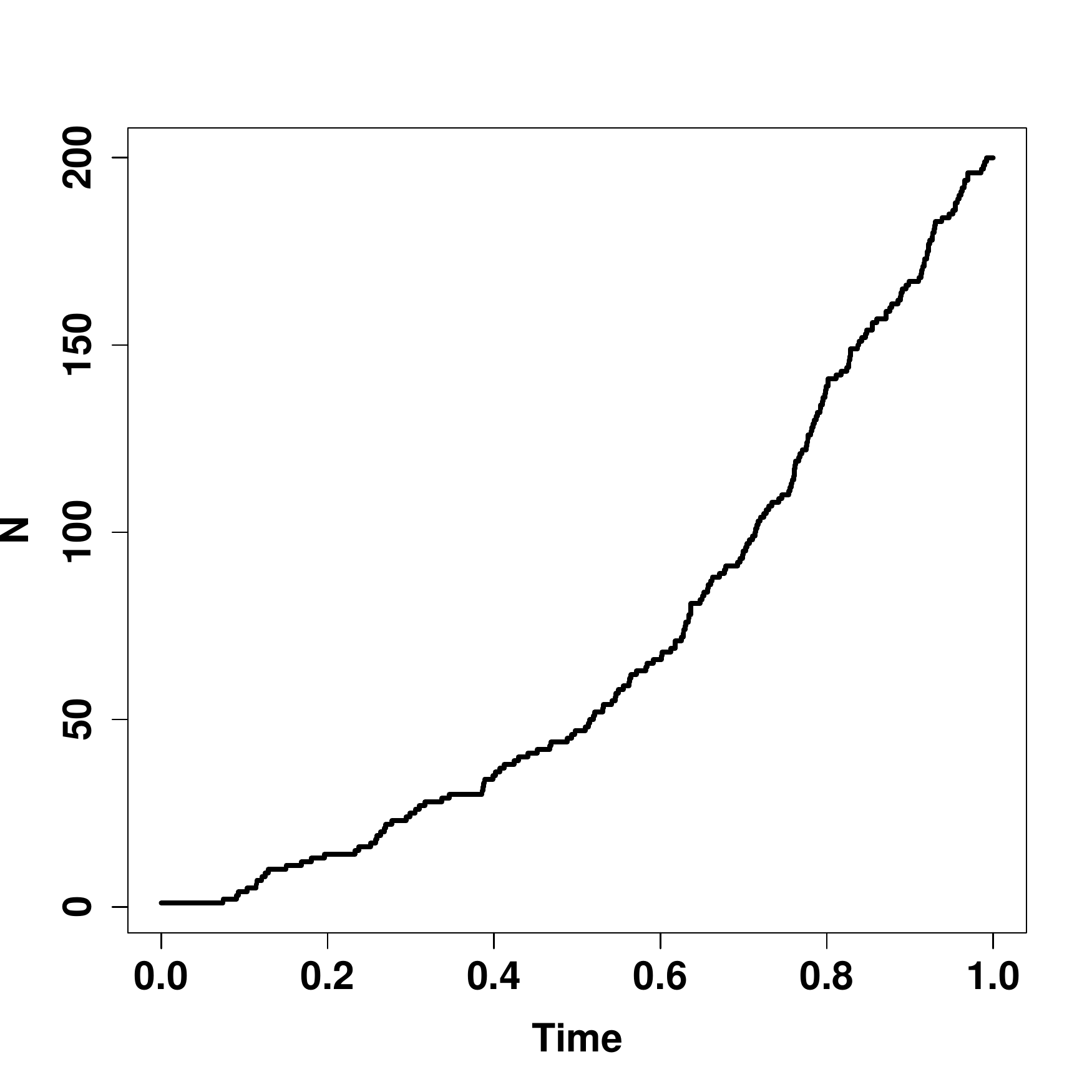} \\
\includegraphics[width=0.242\textwidth,angle=0]{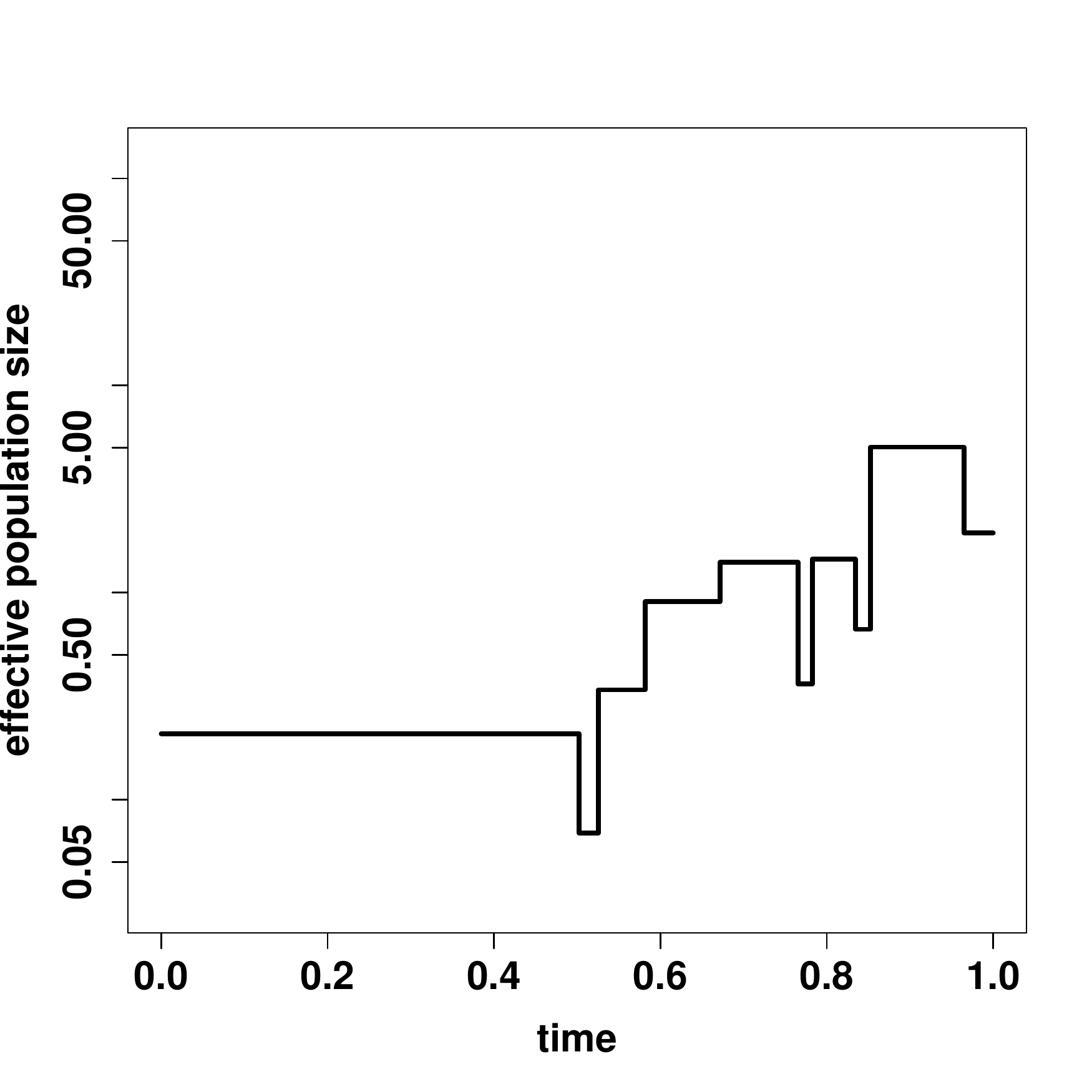}
\includegraphics[width=0.242\textwidth,angle=0]{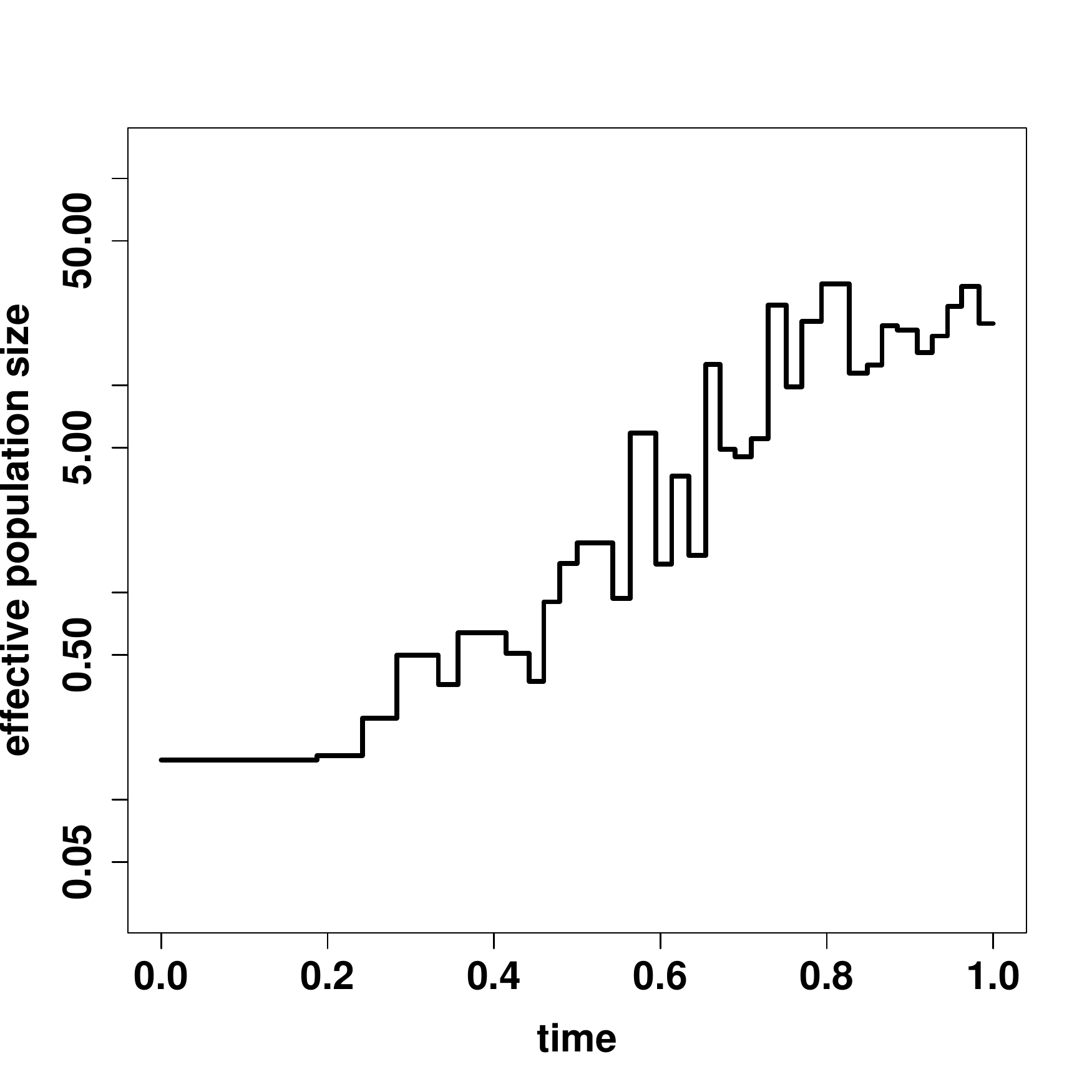} 
\includegraphics[width=0.242\textwidth,angle=0]{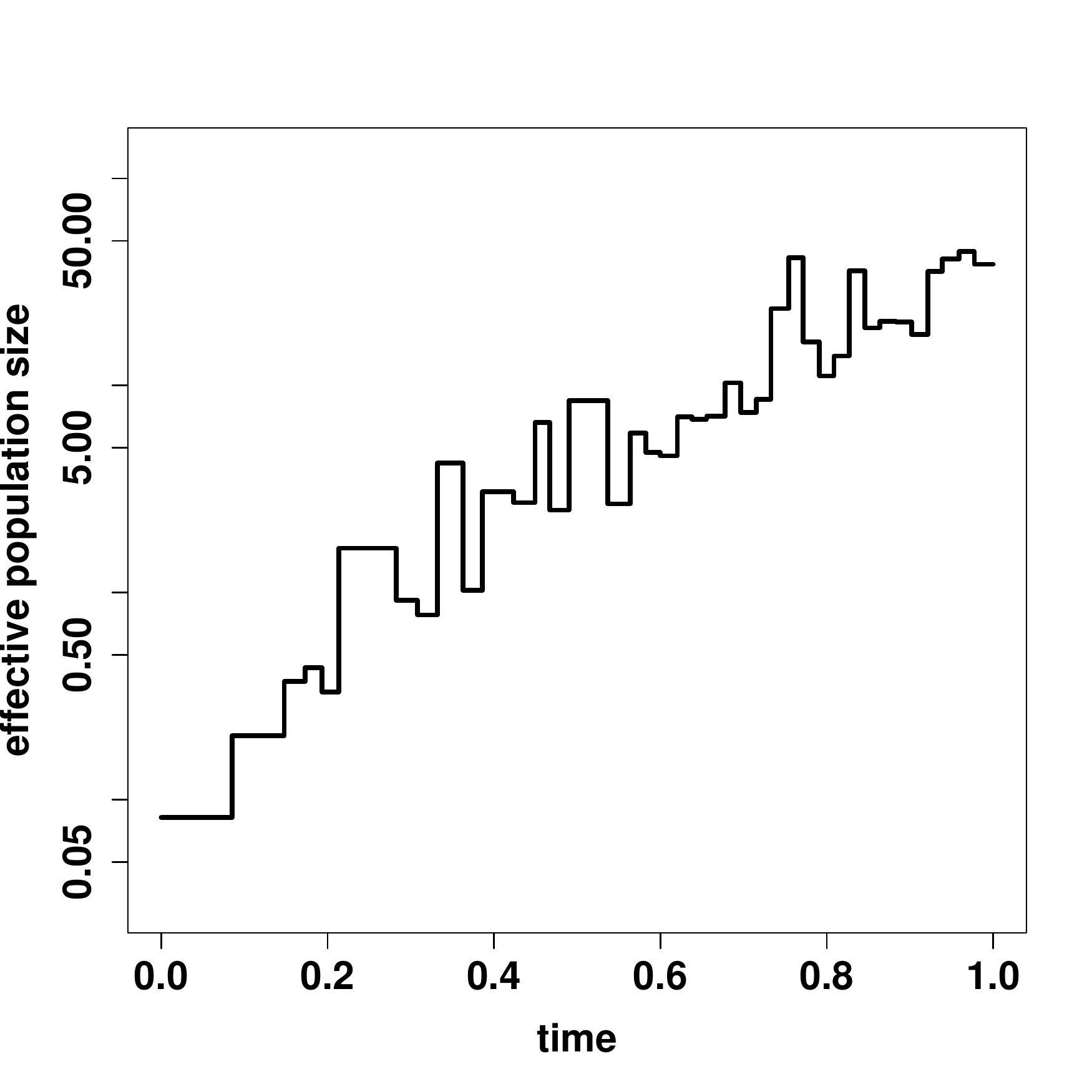}
\includegraphics[width=0.242\textwidth,angle=0]{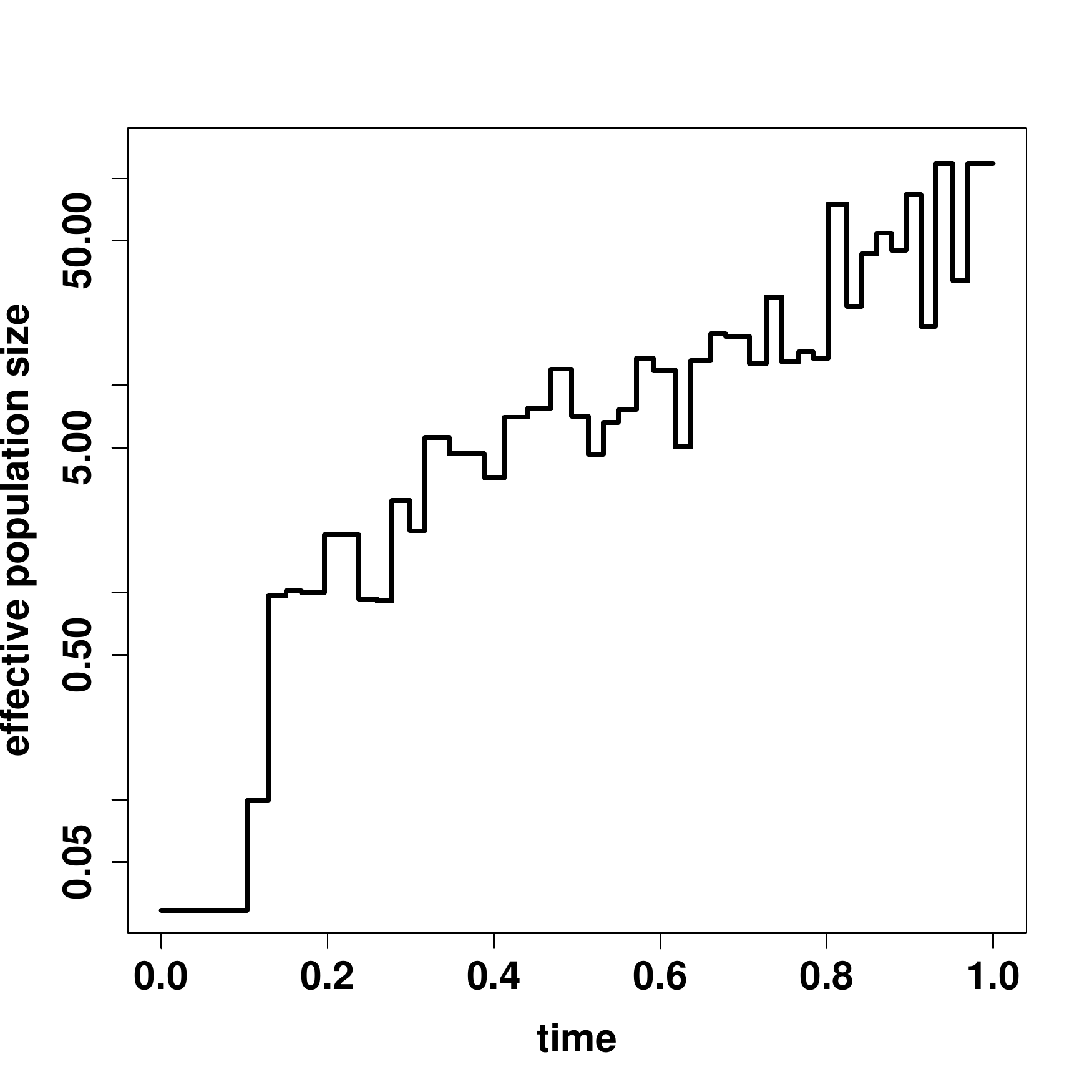} 
}
\end{center}
\caption{
Lineage through time and skyline plots corresponding to the phylogenetic trees
of Fig. \ref{figExampleTDepSpec_1}.
In the first row we provide the lineage through time plots, \pkg{ape}\code{::ltt.plot()},
and in the second row the skyline plots, \pkg{ape}\code{::skyline()} with \code{epsilon=-log(0.95)/3}.
The choice of \code{epsilon} is motivated by how quickly ancestral effects are lost. The parameter
\code{epsilon} controls the temporal structure in the population data \cite{KStrOPyb2001}.
From our perspective we can think of it as controlling how long the coalescent rate
will be approximately constant. The expectation of the OU process after time $t$ equals $e^{-3t}X_{0}+(1-e^{-3t})\theta$.
Hence, we may ask how much time needs to pass for the expectation of the trait to lose a  ``significant''
amount of the ancestral value. If by a ``significant'' amount of loss we take losing $5\%$ of the ancestral value,
then we obtain the time to lose this as $-\log(0.95)/3$ 
(cf. with phylogenetic half--life \cite{THan1997}).
}
\label{figExampleTDepSpec_2}
\end{figure}

\section{The \pkg{pcmabc} \proglang{R} package}\label{secpcmabc}
\subsection{The ABC algorithm}\label{sbsecABCalg}
Obtaining the exact likelihood for a phylogenetic comparative sample is possible only in special cases.
This is essentially only for normal models \cite{PCMBase,RPANDA} 
or discrete state Markov chains
\cite{JFel1973,JFel1981}. Hence, for these types of models a pleiphora of estimation packages are
available 
(e.g. \cite{KBaretal2012,MButAKin2004,JClaGEscGMer2015,RFit2012,EGooJBruCAne2017,THanJPieSOrz2008,LHaretalGEIGER,MManALamHMor2017}). 
However, beyond these types of models development has yet to take off 
(however see \cite{KBar2012,SBlo2016bioRxiv,FBouVDemEConLHarJUye2018,PDucCLeuSSziLHarJEasMSchDWeg2017,MLanJSchMLia2013}). 

The issue at heart is that the likelihood cannot be obtained exactly.
Hence, alternative methods need to be employed, numerically solving an ODE system
\cite{RFit2010,EGolLLanRRee2011,WMadPMidSOtt2007} or ABC (e.g. \cite{FBok2010,DJhw2018arXiv,GSlaetal2012} and
see a review in \cite{NKutHInn2014})

The \pkg{pcmabc} package implements an ABC algorithm that allows for estimation
of parameters of an arbitrary Markov model of trait dependent speciation.
Let $\Theta$ denote the set of model parameters (known ones and those to be estimated)
and $d(\cdot,\cdot)$ the distance between two phylogenetic comparative
data sets (tip data and phylogeny). The general structure of the ABC algorithm 
in the PCM context is also described in \cite{NKutHInn2014}.
\begin{algorithm}[!htp]
\caption{\pkg{pcmabc} ABC algorithm, \code{PCM\_ABC()}}\label{algABCpcmabc}
\begin{algorithmic}[1]
\State{\textbf{input}: A phylogenetic tree 
$\mathcal{T}^{(n)}$ with $n$ tips and phenotypic observations for each tip}
\State{\textbf{output}: Point estimates of parameters with posterior distribution}
\State Set $\Theta_{1}:=\Theta_{initial}$, $\mathcal{P}:=\emptyset$, $\mathcal{R}=\emptyset$, invtotaldist$:=0$, $j=1$
\For {$i=1$ to abcsteps}
\State accept$=$FALSE
\If{tree fixed}
    \State Simulate phenotypic data on tree according to Alg. \ref{algSimData}
    \Comment{Section \ref{sbsecSimul}, \code{simulate\_phenotype\_on\_tree()}}
\Else
    \State Simulate phenotypic data and tree according to Alg. \ref{algSimDataTree}
    \Comment{Section \ref{sbsecSimul}, \code{simulate\_phylproc()}}
\EndIf 
\State $\rho:=d($simulated data,observed data$)$ \Comment{Section \ref{sbsecDistMeas}}
\If{$\rho<\epsilon$}
    \State $\mathcal{P}[j]:=\Theta_{i}$, $\mathcal{R}[j]=\rho$
    \State accept$=$TRUE
    \State invtotaldist$:=$invtotaldist$+1/\rho$
    \State $j++$
\EndIf
\State $\Theta_{i+1}:=$\Call{ParameterProposal}{
$\Theta_{i}$,accept} \Comment{Section \ref{sbsecUpdate}}
\EndFor
\State $$\hat{\Theta}:=\left(\sum_{j}\mathcal{P}[j]/\mathcal{R}[j]\right)/\mathrm{invtotaldist}$$
\State \textbf{return} $(\hat{\Theta},\mathcal{P},\mathcal{R})$
\end{algorithmic}
\end{algorithm}
The parameter point estimate is the inverse distance weighted average.
We take the inverse distance to take into account how close the simulated
sample resembles the original under the given parameter set.

The ABC inference algorithm is invoked as 
\begin{lstlisting}
PCM_ABC(phyltree, phenotypedata, par0, phenotype.model, fbirth, fdeath, X0, step, abcsteps, eps, 
fupdate, typeprintprogress, tree.fixed, dist_method)
\end{lstlisting}
The first three parameters are ``standard'' ones, \code{phyltree} is the phylogeny in \pkg{ape}'s \cite{APE}
\code{phylo} format, \code{phenotype} data is a matrix of trait measurements (rows are the different tips)
and \code{par0} are the initial starting parameters. As described
in Section \ref{sbsecDistMeas} the ABC algorithm's distance function treats the 
phylogeny and trait data separately, hence there is no need for the order of rows to correspond
to the tips' order. 
The initial parameters object, \code{par0} is a list of named lists. The first list corresponds
to the parameters of the phenotypic evolution process, the second to the speciation
dynamics and the third to the extinction dynamics. Inside each list the user can indicate
which parameters are to be optimized over, which treated as fixed and which are to be positive.
If the user would want some further modifications or transformations of the 
parametrization that is optimized over they will need to provide their
own simulation, birth--death and parameter update functions.
The next parameter, \code{phenotype.model} is a user provided function
that models the evolution of the phenotype (see Section \ref{sbsecSimul}),
\code{fbirth} and \code{fdeath} are user provided birth and death
rate functions (see Section \ref{sbsecSimul}), \code{X0} is the root state,
\code{step} is the step size of the simulation (see Section \ref{sbsecSimul}),
\code{abcsteps} is the number of parameter draws of the ABC algorithm,
\code{eps} is the acceptance threshold for the parameters (see Alg. \ref{algABCpcmabc}),
\code{fupdate} is the parameter update function (see Section \ref{sbsecUpdate}),
\code{typeprintprogress} is a parameter that indicates what sort
of summary should be printed out by the inference algorithm during its progress. The user
is free to provide their own function here. Then, the parameter 
\code{tree.fixed} is a logical variable if the phylogeny's branching dynamics
depend on the phenotype or not (see Alg. \ref{algABCpcmabc})
and lastly \code{dist\_method} is a vector of two entries
indicating the distance measure between the trait data and trees respectively
(see Section \ref{sbsecDistMeas}).

\subsection{Simulating the phylogeny and trait(s)}\label{sbsecSimul}
At the core of an ABC method is the ability to simulate data under the model
given parameter proposals. In our situation we have two possibilities to consider.
Firstly the tree is assumed fixed, i.e. the trait value does not affect the branching
dynamics (Alg. \ref{algSimData}), and secondly the branching rates depend on the phenotype
(Alg. \ref{algSimDataTree}).

The first situation is a standard one and we just mention it for completeness' sake. 
At the root one starts at the initial value (user provided). Then, along the branch, of length $t$,
one simulates the phenotype according to the provided simulation procedure (conditional on the candidate parameters).

The user provides their own simulation function defined as
\begin{lstlisting}
f_user_trait_simul<-function(time,params,X0,step)
\end{lstlisting}
where \code{time} is the duration of the simulation, \code{params} is a named list of model
parameters that are interpreted inside the simulation function, \code{X0} is the initial 
trait value and \code{step} is the simulation's step size.
The output of the function is to be the trajectory of the trait starting from \code{X0}
for time \code{time} with values at points $i\cdot$\code{step}.
The output object is a matrix. The first row are the time instances (starting from $0$, in steps
of size \code{step}, correction for the actual time from the root is done later by the package),
and the next rows the trait values at the time instances. It is important to point out that the trait
can be of any dimension.

Special support is provided for trait simulation by the \pkg{yuima} package \cite{ABroetalYUIMA2014} ,
through the \code{simulate\_sde\_on\_branch()} function.
The call is
\begin{lstlisting}
simulate_sde_on_branch(branch.length, model.yuima, X0, step)
\end{lstlisting}
where \code{model.yuima} replaces \code{params} and defines the stochastic differential
equation that should be used to model the trait. The \code{model.yuima} parameter
is the output of the function \code{yuima::setModel()}. Details how to 
construct it can be found in \code{yuima::setModel()}'s manual pages and also in
\code{simulate\_sde\_on\_branch()}'s manual page there is an example how to set it.

After simulating along the branch, a speciation event occurs. Along all the daughter
lineages independent evolution is repeated as above. The starting condition for these
is the value at the parental node. More concisely we describe this in Alg. \ref{algSimData}.
\begin{algorithm}[!htp]
\caption{Trait simulation algorithm, \code{simulate\_phenotype\_on\_tree()}}\label{algSimData}
\begin{algorithmic}[1]
\State{\textbf{input}: A phylogenetic tree 
$\mathcal{T}^{(n)}$ with $n$ tips, a Markovian model of trait evolution and trait value at root, $x_{0}$}
\State{\textbf{output}: Simulated values of the trait along the whole tree}
\Procedure{simulate\_on\_tree}{tree, simulation\_model, 
$x_{0}$}
\State $X_{\mathcal{T}^{(n)}}:=list()$
\For{daughter branch $i$ of root of tree}
\State $X_{i}:=$simulate trajectory on $i$--th branch
\State $X_{\mathcal{T}^{(n)}}[i]:=$join$(X_{i},$\Call{simulate\_on\_tree}{subtree
from $i$--th daughter node, simulation\_model, $X_{i}[$end$]$}$)$ 
\EndFor
\State \textbf{return} joined entries of $X_{\mathcal{T}^{(n)}}$
\EndProcedure
\State \textbf{return} \Call{simulate\_on\_tree}{
$\mathcal{T}^{(n)}$, simulation\_model, $x_{0}$}
\end{algorithmic}
\end{algorithm}
Algorithm \ref{algSimData} is invoked by calling
\begin{lstlisting}
simulate_phenotype_on_tree(phyltree, fsimulphenotype, simul.params, X0, step)
\end{lstlisting}
\code{phyltree} is the phylogeny, \code{X0} the root state, \code{step} the step size.
The function to simulate the phenotype is passed through \code{fsimulphenotype}
and the named list of parameters passed to it through \code{simul.params}. To simulate the
phenotype through the \code{simulate\_sde\_on\_branch()} function one sets
\code{fsimulphenotype="sde.yuima"}.

The second situation is more involved. Branching dynamics are trait dependent
so a straightforward simulation, by time steps of some size, would be very computationally heavy.
Hence, we employ a variation of the rejection sampling algorithm for the Inhomogeneous Poisson process 
(Proposition p. $32$, \cite{SRos2006}). 
\begin{algorithm}[!htp]
\caption{Trait and tree simulation algorithm,\code{simulate\_phylproc()}}\label{algSimDataTree}
\begin{algorithmic}[1]
\State{\textbf{input}: Tree height, number of tips, 
a Markovian model of trait evolution and trait value at root, $x_{0}$}
\State{\textbf{output}: Tree and simulated values of the trait along the whole tree}
\Procedure{simulate\_on\_tree}{height, simulation\_model, 
$x_{0}$}
\State $X(t):=$simulate trait trajectory on lineage with length height
\State Calculate the birth rate $\lambda(t)$ as a function of $X(t)$
\State Calculate the death rate $\mu(t)$ as a function of $X(t)$
\State $\Lambda=\max  \lambda(t)$
\State Decompose $\lambda(t)=\Lambda p_{\lambda}(t)$
\State Simulate a Poisson process for time height and rate $\Lambda$
\State Accept events from the Poisson process with probability $p_{\lambda}(t)$ \Comment{these will be branching events}
\State $\mathcal{M}=\max  \mu(t)$
\State Decompose $\mu(t)=\mathcal{M} p_{\mu}(t)$
\State Simulate a Poisson process for time height and rate $\mathcal{M}$
\State Accept events from the Poisson process with probability $p_{\mu}(t)$ \Comment{these will be extinction events}
\State Check if a death event is along the lineage. If so remove all branching events following it.
\For{each accepted branching event $i$}
\State $\mathcal{T}_{i}:=$\Call{simulate\_on\_tree}{height decreased by time of node $i$, simulation\_model, 
trait value at node $i$}
\EndFor
\State \textbf{return} joined $\mathcal{T}_{i}$ and $X(t)$
\EndProcedure
\State \textbf{return} \Call{simulate\_on\_tree}{
Tree height, simulation\_model, $x_{0}$}
\end{algorithmic}
\end{algorithm}
Inside Alg. \ref{algSimDataTree} we did not write how one uses the number of tips. 
This is a very technical detail in the implementation. The simulation will terminate if it reaches the maximum
number of tips (contemporary or also including extinct, the user decides which to provide). 
However, there is no guarantee that a tree
with this number of tips will actually be simulated. If there are death events, then the process
may die out before the value is reached. Alternatively, too few birth events get simulated and 
again the desired number is not reached.

The simulation function is more involved, then in the case of simulation on a fixed tree
\begin{lstlisting}
simulate_phylproc(tree.height, simul.params, X0, fbirth, fdeath, fbirth.params, fdeath.params, fsimulphenotype, n.contemporary, n.tips.total,step)
\end{lstlisting}
Some parameters are self explanatory, \code{tree.height} is the height of the tree,
\code{X0} the root state, \code{step} the simulations step size, \code{n.contemporary}
the number of contemporary tips, \code{n.tips.total} total number of tips,
including extinct ones. As before \code{fsimulphenotype} is the function to simulate the trait
but if it equals \code{"sde.yuima"}, then \code{simulate\_sde\_on\_branch()} is invoked. 
Parameters of trait simulation function are passed as a named list in \code{simul.params}.

The user additionally provides the birth and death rates (the latter can be passed as \code{NULL} meaning
that it is a pure--birth process) as functions. The birth rate function is provided through
the \code{fbirth} parameter and the death function through the \code{fdeath} parameter.
The parameters of the two rate functions are passed through the named lists 
\code{fbirth.params} and \code{fdeath.params}. Both can be \code{NULL}.
It is up to the user to make sure that the functions
return positive values. Some rate functions are provided inside the package, namely \code{"rate\_const"}
(constant rate, with a possibility to switch to a different value if the first trait variable
exceeds some value) and \code{"rate\_id"} (equals the value of the first trait or linear transformation of it, 
see manual for options). 

The first parameter of the birth and death function has to be the trait vector. The user should
remember that the first entry of the trait vector is time (counted from the start of the lineage, i.e.
from the speciation event where the lineage appeared). Hence, some sort of time--heterogeneity is possible.
The second parameter is the named list of rate function parameters. Both rate functions should return a
single non--negative real number. 

It is worth pointing out that the package has also basic graphic capabilities. The
function \code{draw\_phylproc()} takes the output of the \code{simulate\_phylproc()}
function and draws the trajectory of the trait (as can be seen in Fig. \ref{figExampleTDepSpec_1}).
If one wants the tree (in \code{phylo} format), then one can access it through
the field \code{tree} of \code{simulate\_phylproc()} output. It is good to know that the
tree has a couple of extra fields with respect to the usual fields of the \code{phylo} format tree. 
In particular it has the field \code{tree.height} which stores the height of the tree (time from
origin to contemporary tips) and \code{node.heights} which stores the time from each node to the origin  
of the tree.

\subsection{The summary statistics and distance measure}\label{sbsecDistMeas}
A key element of an ABC algorithm is the choice of the summary statistic for the simulated/observed
sample and the distance function between the observed and simulated statistics. 
In our situation the sample consists of two components---the phylogeny and observed trait values. 

The \pkg{pcmabc} package offers various possibilities in this respect.
We describe the ones that through numerical experiments seemed to work best.
It turned out that looking at the sample mean and variance of the trait
values was the best option. Statistics based on tips' means and variances have already been considered 
in the PCM setting \cite{FBok2010,DJhw2018arXiv,GSlaetal2012}. 
However, the situation is different in the previous two cases. 

In \cite{GSlaetal2012} they consider terminal lineages corresponding to higher taxonomic levels.
Each such tip, contains a number of species for which phylogenetic relationships might not be resolved.
Then, the differences between the means and variances of the trait measurements from species
inside each higher order tip are taken. 
Earlier, in a similar spirit just the variances inside clades were compared \cite{FBok2010}.
Similarly in \cite{DJhw2018arXiv} the mean and variance of the difference between tip measurements
are calculated. 

Our situation is different as we do not consider a fixed tree, but a random one so we will not 
have a correspondence between the tips. On the other hand even with a fixed tree there are
multiple possible symmetries, from the perspective of the trait process, in the tip labellings.
Take for example the simplest case of a cherry (i.e. two tips stemming from a single ancestral
node). Then, as trait evolution is independent following speciation, one cannot distinguish 
between the two tips, unless there is some additional information, like branch
specific parameters. Without such taking the difference between tips by the original
labels might not be the optimal choice.

The distance between trees is actually more involved. It seems that the weighted and normalized
Robinson--Foulds metric \cite{DRobLFou1979,DRobLFou1981} 
implemented in the \pkg{phangorn} \cite{phangorn} package as \code{wRF.dist()} with \code{normalize=TRUE}
seemed to work the most effectively in our experiments. These are the 
default distance functions. However, it should be pointed out that the
simulation--based tests were performed using OU models of trait evolution. These
are normal processes and hence all information is stored in the mean and variance. Should
the trait evolve as a non--normal process other distance measures could be more appropriate.

\subsection{Proposing parameters}\label{sbsecUpdate}
The standard ABC algorithm draws parameter proposals from the prior
distribution (see ABC steps description in \cite{NKutHInn2014}). 
In our situation, due to the complexity of the sample, this would have resulted 
in nearly all parameter proposals being rejected. Hence, we employed
a hybrid proposal algorithm that attempts to explore in detail the parameter
space around ``good'' proposals. 

If the previous parameter set was rejected, the inference algorithm samples
each parameter uniformly from the interval $[-10,10]$. Such a restriction
was chosen not to have extreme parameter values. If this restriction
is problematic, a user may provide their own proposal function as described below.
If the previous parameter set was accepted, then each new parameter is the previous
parameter modified by a mean--zero normal deviation (with user specified standard deviation).

The user is allowed to specify which parameters are to be positive
(transformation by taking the exponential). 
As a result we cannot say that we obtain a sample from the posterior distribution
as an usual ABC algorithm would result in. However, this method
seems to result in decent parameter estimates as presented in Section \ref{secSimul}.

However, the user is also free to specify their own parameter update function.
The function has to handle the call
\begin{lstlisting}
f_user_update(par,par0,baccept,ABCdist,phenotypedata,phyltree)
\end{lstlisting}
where \code{par} is the list (as described in Section \ref{sbsecABCalg})
of parameters from the previous step, \code{par0} are the initial
parameters provided to \code{PCM\_ABC()} (see Section \ref{sbsecABCalg}),
\code{baccept} is a logical variable indicating if the \code{par} parameters
were accepted (\code{TRUE}) or not (\code{FALSE}),
\code{ABCdist} is the distance between the observed
and simulated (under \code{par}) data, \code{phenotypedata} is 
the original data matrix of trait measurements and \code{phyltree} is the original phylogenetic tree.

In particular this means that it is possible to change the inference to a usual
ABC algorithm with independent proposals from the prior. The user defined
function just samples from the prior ignoring all information on the previously considered
parameter set.

\section{Proof of concept simulation study}\label{secSimul}

We performed a simple simulation study to evaluate whether
the ABC inference package can capture any signal on the 
trait--dependent speciation process, based only on the 
contemporary sample and phylogeny. This is not a trivial question,
as it is not clear what exactly is estimable in such a setup. 
The trait and branching dynamics interact with each other with
many potential masking effects. In the PCM context, such
masking effects occur in even simpler setups
(e.g. \cite{KBarSGleIKajMLas2017}). Hence, we aim at a proof of concept
study to evaluate the potential usefulness of the inference algorithm.

We simulate a univariate trait that follows a OU process, defined as
$$
\ud X_{t} = -\alpha(X_{t}-\psi)\ud t + \sigma \ud B_{t},
$$
\begin{lstlisting}
simulate_OU_sde<-function(time,params,X0,step){
    A <- c(paste("(-",params$a11,")*(x1-(",params$psi1,"))",sep=""))
    S <- matrix( params$s11, 1,1)
    yuima.1d <- yuima::setModel(drift = A, diffusion = S, state.variable=c("x1"),solve.variable=c("x1") )
    simulate_sde_on_branch(time,yuima.1d,X0,step)
}
\end{lstlisting}
with parameters
\begin{lstlisting}
true_sde.params<-list(a11=1,s11=1,psi1=0, positivevars=c(TRUE,TRUE,FALSE), abcstepsd=rep(0.5,3))
\end{lstlisting}
The reader can also see how one indicates positive parameters \code{positivevars} and the standard deviation
for the Gaussian update of proposed parameters \code{abcstepsd}.
We assume a pure birth process, with speciation rate function
\begin{lstlisting}
fbirth_rate_constrained<-function(x,params,...){
    x<-x[2]
    params$scale/(1+exp(-x))
}
\end{lstlisting}
with parameters
\begin{lstlisting}
true_birth.params<-list(scale=5,abcstepsd=0.5,positivevars=c(TRUE,TRUE),fixed=c(FALSE))
\end{lstlisting}
We chose the OU model for the phenotype because it is the current gold standard in PCMs
for modelling adaptive evolution. Its dynamics are very well understood---after 
an initial ``burn--in period'' the trait will stabilize around its stationary
distribution. Furthermore, for a constant rate birth--death, process 
``memory effects'' have been intensively studied (e.g. \cite{RAdaPMil2015,CAneLHoSRoc2017,KBarSSag2015a}).
Hence, in our setup we should expect the birth rate to oscillate in a controlled manner, after
each lineage reaches its stationary distribution. However, as the \code{scale=5} parameter is significantly
greater than $\alpha=1$ we are in the fast branching (or slow adaptation) regime. This induces
strong correlations (through remembered ancestral effects) between evolving lineages, making
estimation more difficult and in a way introducing less straightforward process dynamics. 

The speciation rate takes values between $(0,$\code{scale}$)$. With negative trait values it decreases to $0$,
with positive increases up to \code{scale}. As the trait follows an OU process with 
stationary mean $0$, it will oscillate around $0$. Hence, the birth rate should oscillate around \code{scale}$/2$.
The main question of interest is if the inference procedure will be able to identify the value
of \code{scale} and also of the OU process's parameters.
The reader can also see how one indicates parameters which are not to be optimized over. 
The logical values in the field \code{fixed} tell this to the inference procedure. If it were
\code{TRUE}, then \code{scale} would never be changed from its initial value. 

We follow a simple procedure of simulating data under the model and then calling \code{PCM\_ABC()}
to recover parameters given the simulated phylogeny and contemporary trait measurements.
We take \code{abcsteps=1000}, number of tips of the phylogeny $200$, simulation step size $0.001$,
the distance between phylogenies is taken as the  Robinson--Foulds metric and the distance between
the traits compares the sample mean and variances. We take \code{eps=0.2} as the parameter
acceptance cut--off when comparing the summary statistics. 
Furthermore, we used the default setting that the inference algorithm samples
each new parameter proposal after rejection, uniformly from the interval $[-10,10]$.
This however is on the scale used in estimation, so if a parameter is assumed positive
($\alpha$, $\sigma$, \code{scale}) then $[-10,10]$ are bounds on the log scale. On the
real scale these translate to $[e^{-10},e^{10}]$. Furthermore,
we bounded the \code{scale} parameter (on its actual scale) by $10$, i.e.
\code{scale}$\in [e^{-10},10]$.

For the fixed tree simulation part we did not use \pkg{pcmabc}'s functionality. This was 
to avoid any potential bias and see how \pkg{pcmabc} can recover parameters from a completely
independent of its code simulation. We simulated the phylogeny using the \pkg{TreeSim} 
package
\begin{lstlisting}
phyltree<-TreeSim::sim.bd.taxa(numbsim=1,n=200,lambda=1,mu=0)[[1]]
\end{lstlisting}
and then the OU--evolving traits using the \pkg{mvSLOUCH} package
\begin{lstlisting}
OUOUparameters<-list(vY0=matrix(0,1,1),A=matrix(1,1,1),mPsi=matrix(0,1,1),Syy=matrix(1,1,1))
OUOUdata<-mvSLOUCH::simulOUCHProcPhylTree(phyltree,OUOUparameters)
\end{lstlisting}

\begin{figure}[!ht]
\begin{center}
{%
\includegraphics[width=0.45\textwidth]{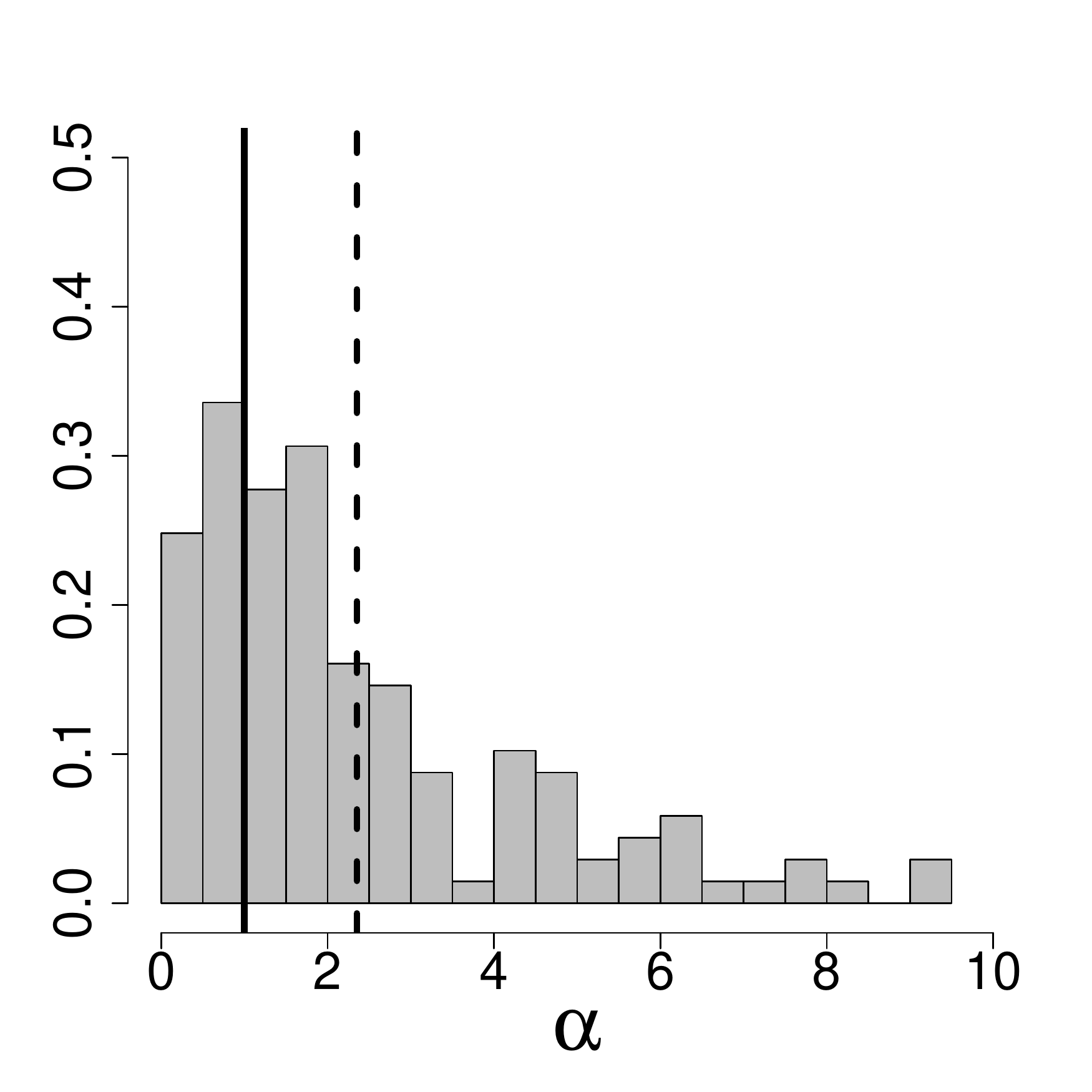}
\includegraphics[width=0.45\textwidth]{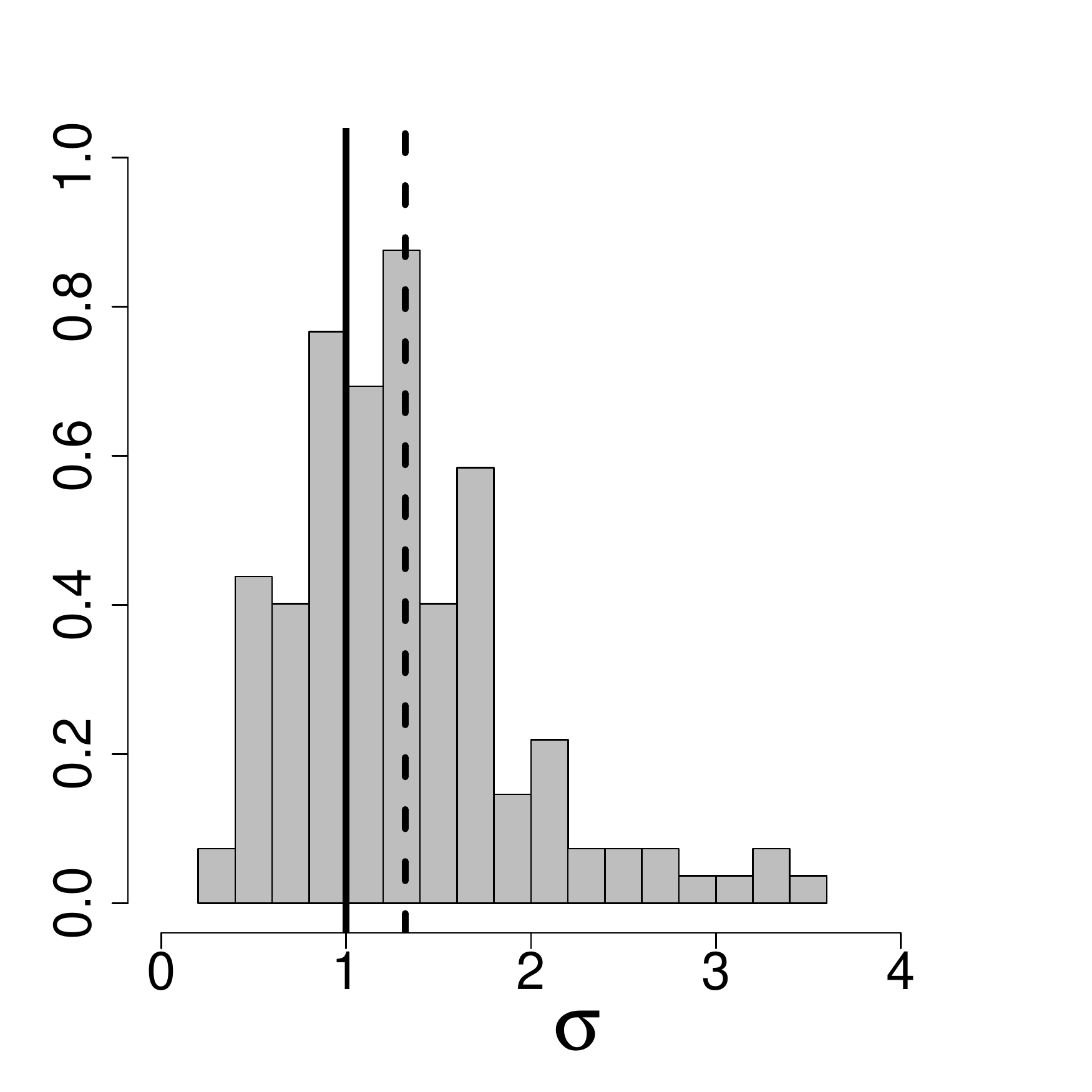} \\
\includegraphics[width=0.45\textwidth]{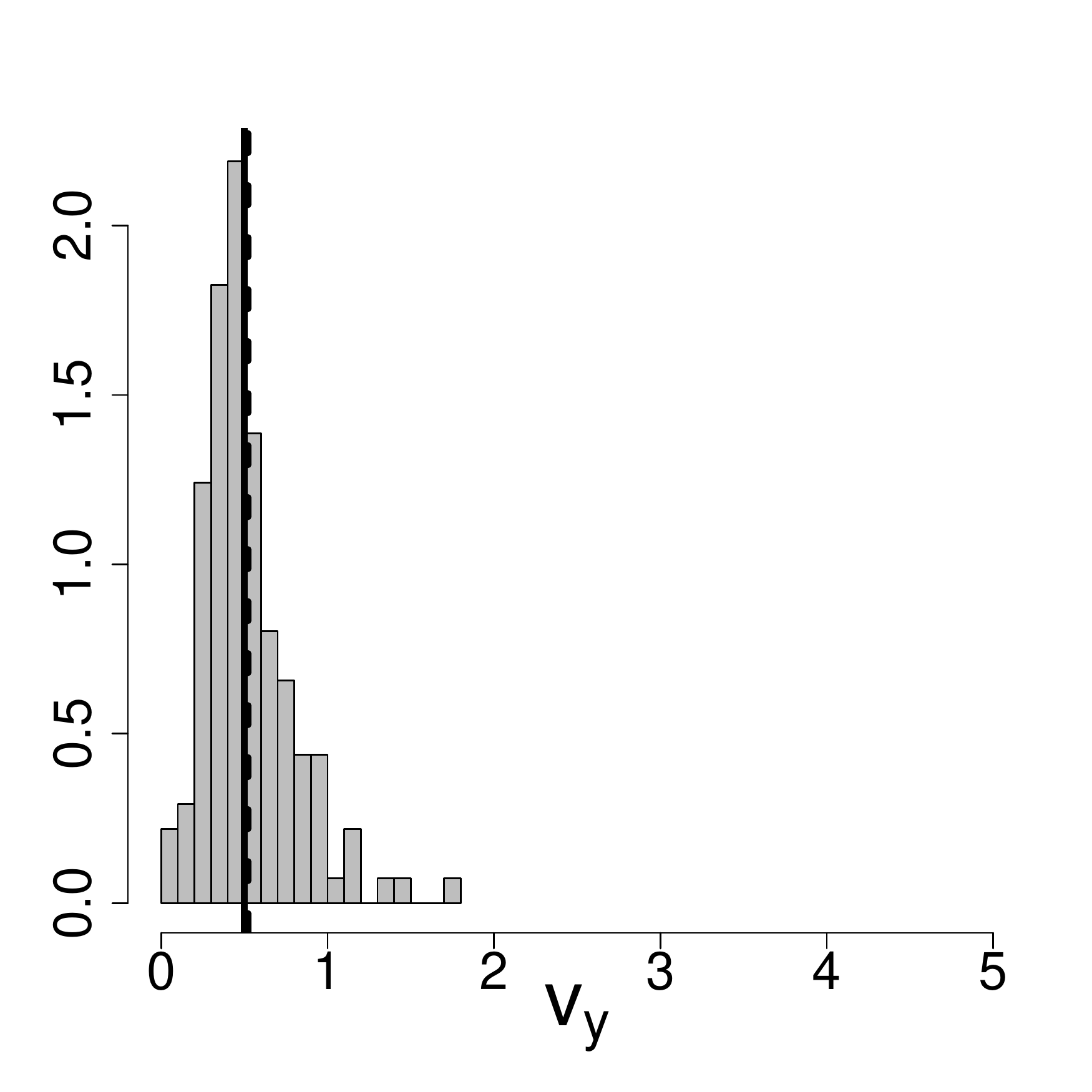}
\includegraphics[width=0.45\textwidth]{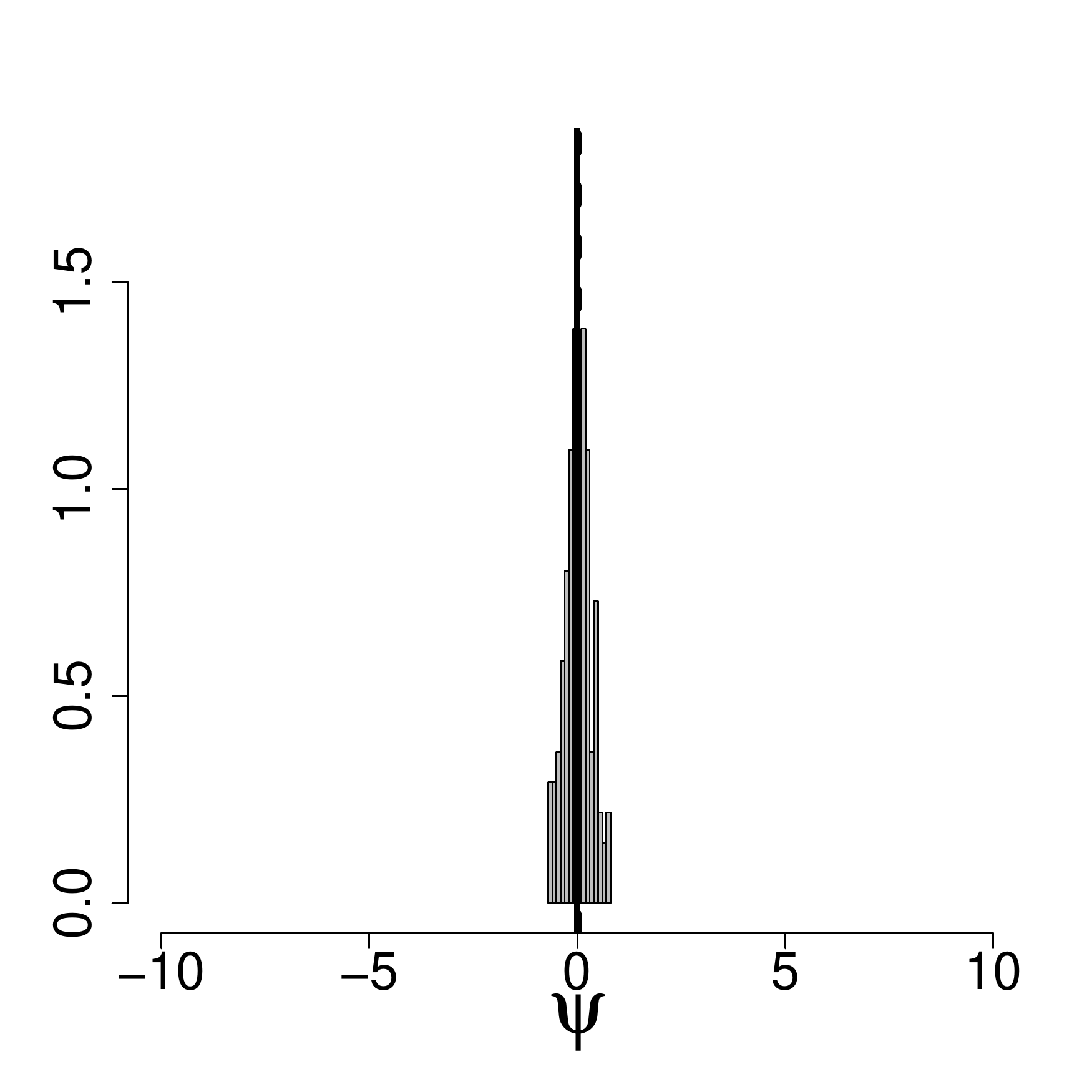}
}
\end{center}
\caption{
Histograms of the parameter estimates in the fixed tree case, top left: $\alpha$, top right $\sigma$, bottom left: $\psi$, bottom right: $v_{y}$.
The solid gray line is the true value, the dashed the mean of the estimates.}
\label{figHistsABCfixedRes}
\end{figure}

\begin{figure}[!ht]
\begin{center}
{%
\includegraphics[width=0.32\textwidth]{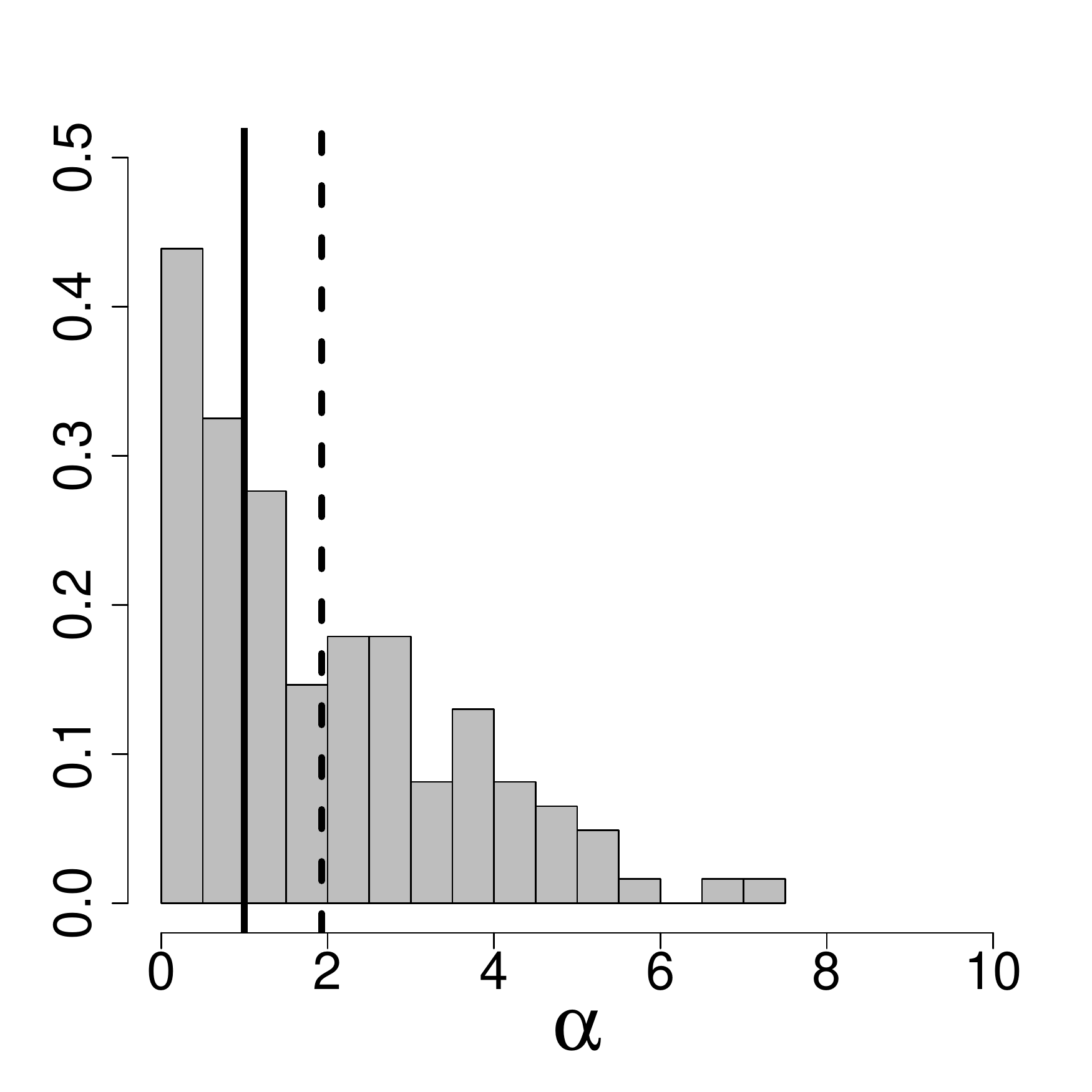}
\includegraphics[width=0.32\textwidth]{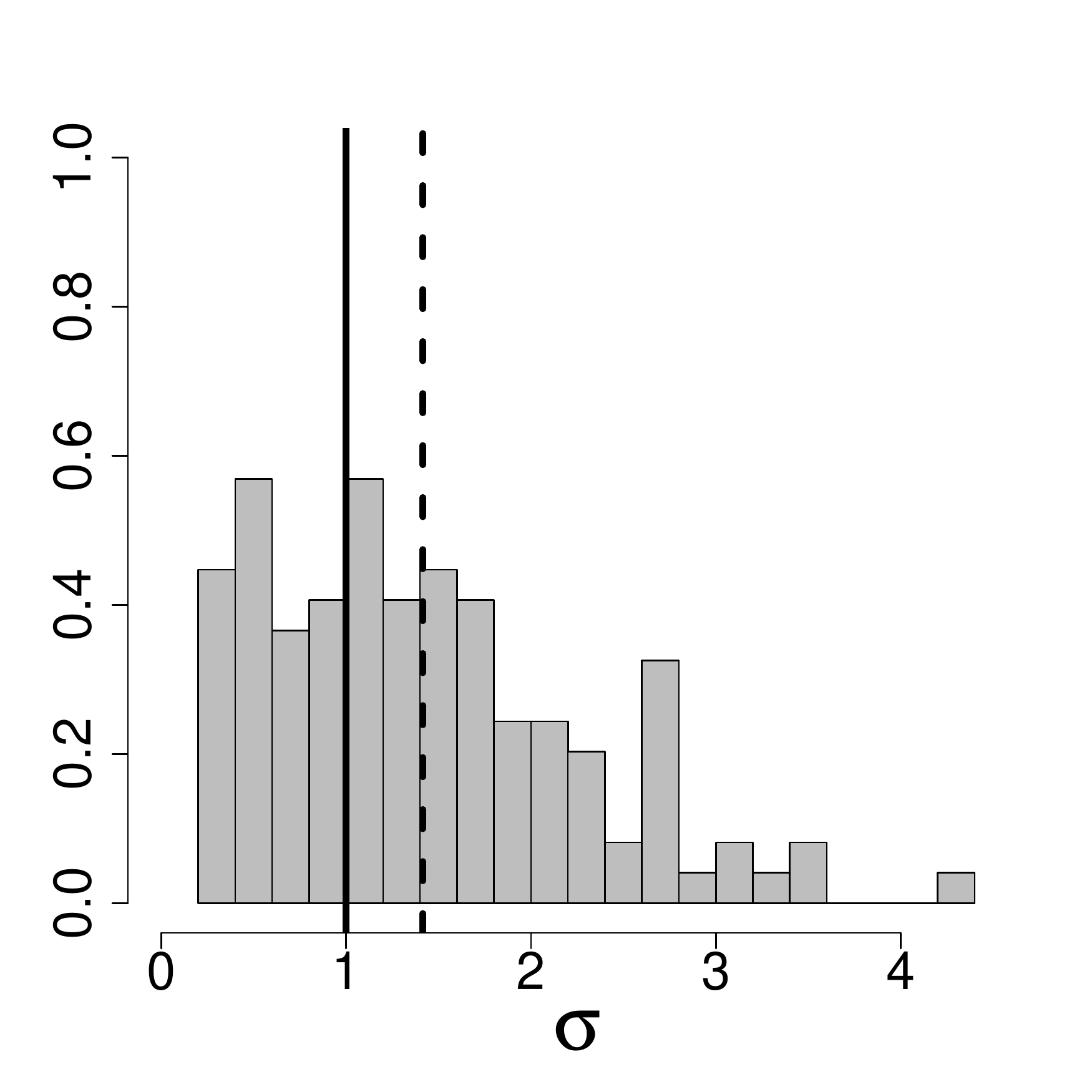} 
\includegraphics[width=0.32\textwidth]{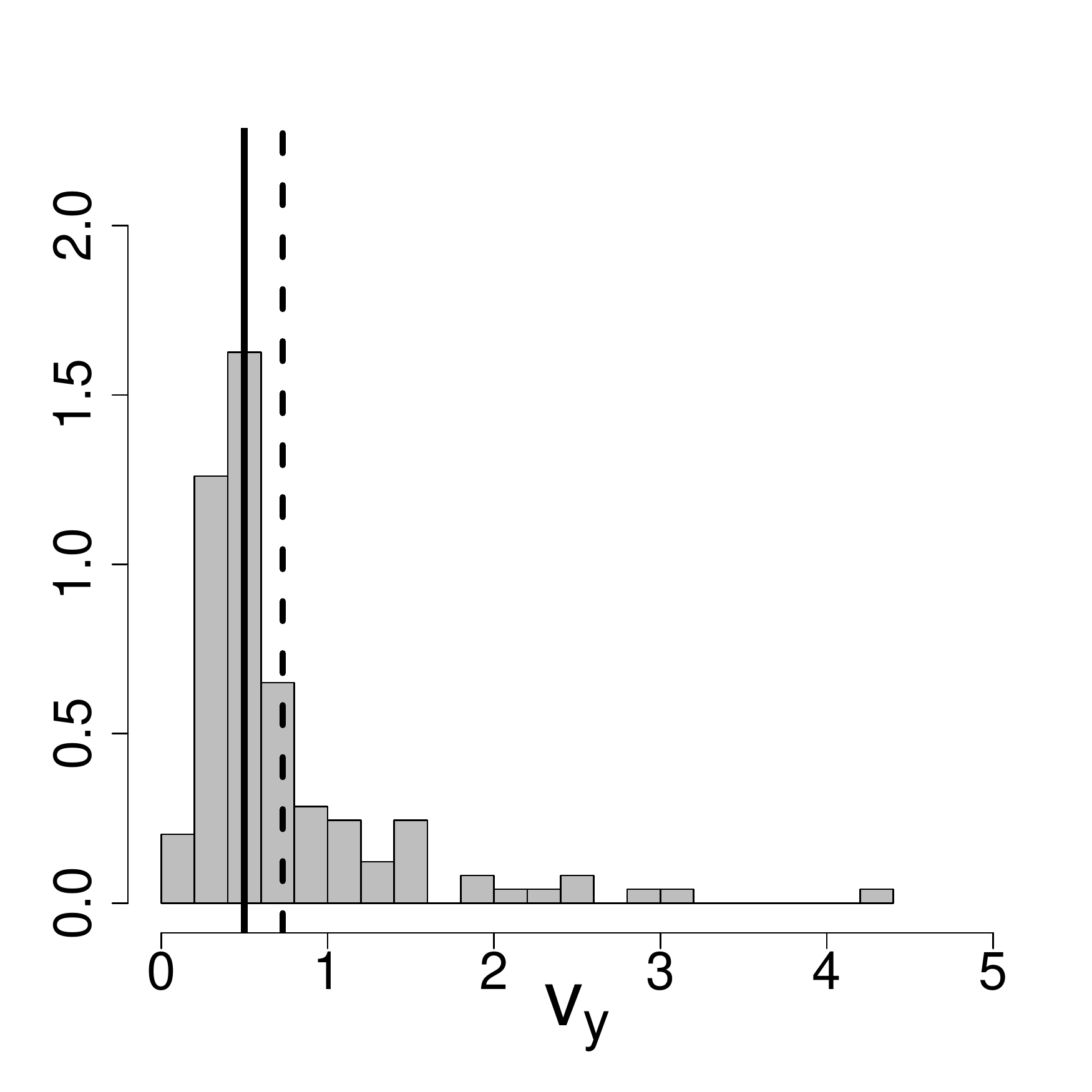} \\
\includegraphics[width=0.32\textwidth]{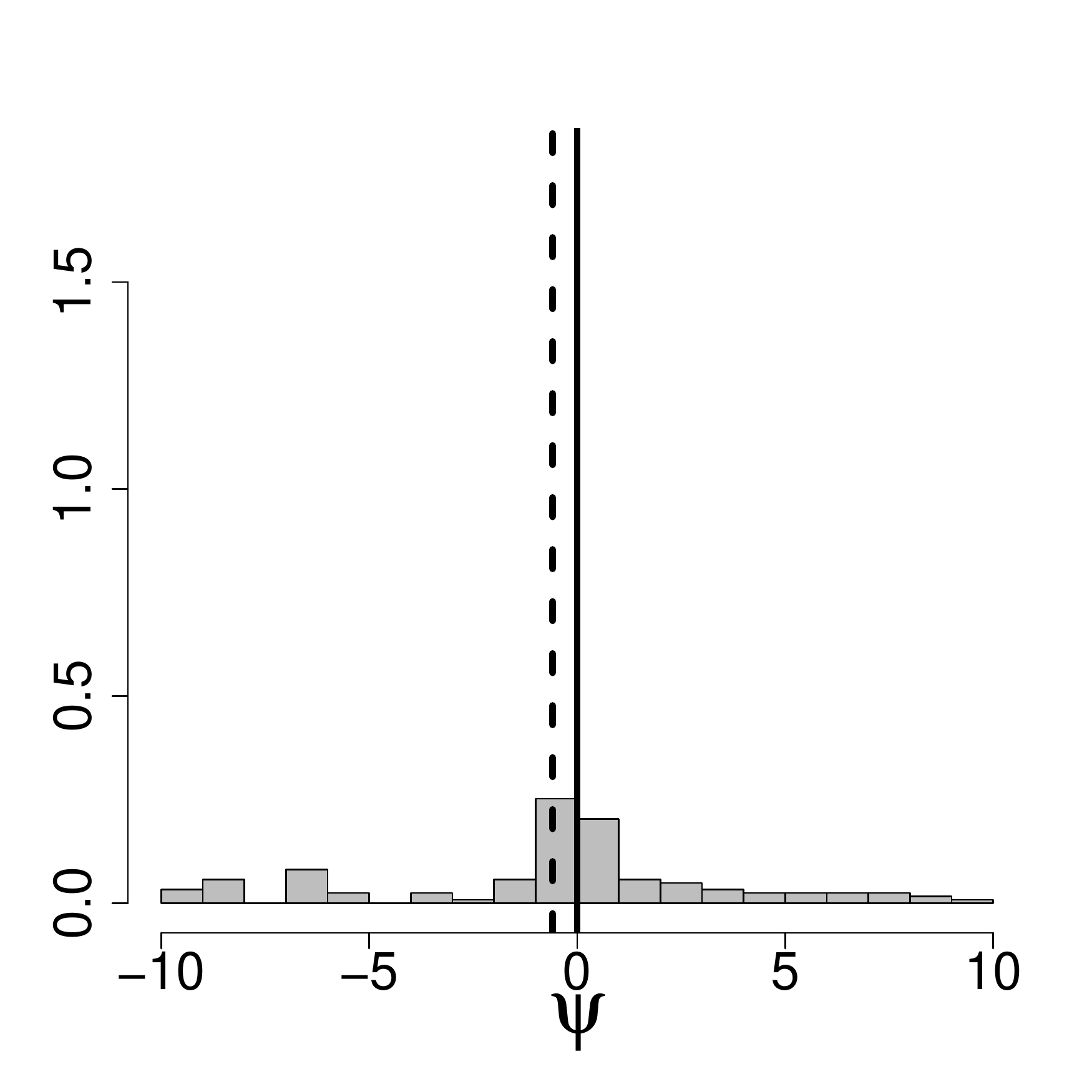}
\includegraphics[width=0.32\textwidth]{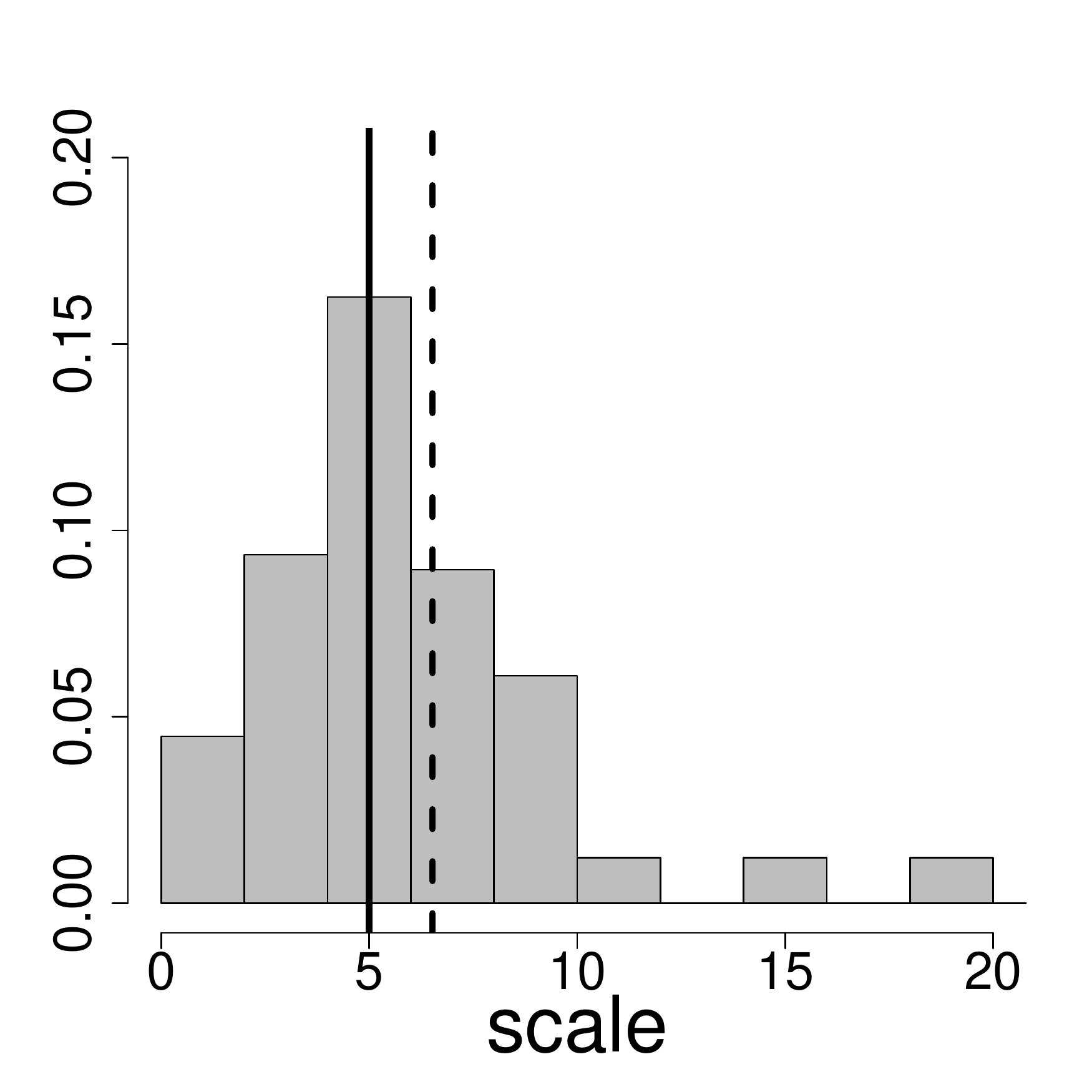}
}
\end{center}
\caption{
Histograms of the parameter estimates in the fixed tree case, top left: $\alpha$, top center: $\sigma$, top right: $v_{y}$,
bottom left: $\psi$, bottom right: \code{scale}.
The solid gray line is the true value, the dashed the mean of the estimates.
}
\label{figHistsABCnonfixedRes}
\end{figure}

We perform $170$ repeats of the simulate and recover parameters procedure.
We recover parameters for the situation where the tree is assumed fixed (here only
OU parameters can be recovered) and where the tree is assumed to be non--fixed (we can
also recover \code{scale}). We present the summary of the results of the simulation
study in Tab. \ref{tabABCRes} 
and in Figs. \ref{figHistsABCfixedRes},  \ref{figHistsABCnonfixedRes} the histograms of the estimated values.

All simulations were done in \proglang{R} version $3.4.2$ running on an openSUSE $42.3$ (x$86$\_$64$) box.
A major handicap of the study were the long running times. About two weeks were required to obtain
$20$ repeats of the simulation--estimation procedure (with $1000$ parameter proposals).

\begin{table}[htb]
\begin{center}
\begin{tabular}{cc|cccc}
&& \multicolumn{2}{c}{fixed tree} & \multicolumn{2}{c}{not fixed tree} \\
parameter & true value & mean & sd & mean & sd \\
\hline
$\alpha$ & $1$ & $2.352$ & $2.069$ & $1.928$ & $1.601$\\
$\sigma$ & $1$ & $1.319$ & $0.626$ & $1.415$ & $0.842$ \\
$v_{y}$ & $0.5$ & $0.521$ &  $0.270$ & $0.731$ & $0.647$ \\
$\psi$ & $0$ & $0.027$ & $0.311$ & $-0.589$ & $4.233$\\
\code{scale} & $5$ & --- & --- & $6.523$ & $5.567$ \\
\hline
\end{tabular}
\caption{
Mean and variance of parameter estimates for $170$ repeats of the simulate and re--estimate
procedure. It is important to notice that for the fixed tree case $33$ of the repeats
resulted in an error (and no estimate) while for the non--fixed tree case $47$.
The sample's standard deviation is abbreviated as sd. We also present
estimates of the composite parameter $v_{y}=\sigma^{2}/(2\alpha)$, the stationary
variance of the OU process.
}\label{tabABCRes}
\end{center}
\end{table}

The results of the simulation--estimation are on the one hand not surprising. The 
parameters $\alpha$, $\sigma$ are not easy to estimate, as was indicated in \cite{CCreMButAKin2015}.
The parameters  $\psi$, $v_{y}$ (OU's stationary mean and variance) are much easier to estimate, as they should be 
close to the sample average and variance respectively 
(when the tree is fixed, e.g. \cite{KBarSSag2015a}).

On the other hand, what is optimistic in the fixed tree case is that these parameter estimates
are close to the true ones. Previous studies were performed for likelihood--based inference 
methods. Our ABC approach does not use the likelihood and hence, is at a serious disadvantage.
Furthermore, only $1000$ ABC steps are done (including the rejected proposals), due
to running times. Despite this, the estimated parameter values are in a reasonable range.

When one looks at the more interesting situation, where the tree is not taken to be fixed, 
one can also be optimistic. The key parameter, \code{scale} seems estimated not too
far away from its true value. Furthermore, a slight improvement in the value 
of $\alpha$ is visible. The marked deterioration of the $\psi$ parameter is surprising
and warrants a more in--depth analysis. One possible direction of further development
is to develop (as the interface allows this) hybrid ways of parameter proposals.
For example for $\psi$ and $v_{y}$ the sample mean and variance values
could be used more directly.

\section{\pkg{pcmabc}: an \pkg{R} tutorial}\label{secTutorial}
One begins work with the package by loading it
\begin{lstlisting}
library(pcmabc)
\end{lstlisting}
Then, one needs to read--in the trait observations and tree. As this is project specific we will
rather simulate it using \pkg{pcmabc}'s capabilities.
We first define the trait simulation function (the same OU process as in Section \ref{secSimul})
\begin{lstlisting}
simulate_OU_sde<-function(time,params,X0,step){
    A <- c(paste("(-",params$a11,")*(x1-(",params$psi1,"))",sep=""))
    S <- matrix( params$s11, 1,1)
    yuima.1d <- yuima::setModel(drift = A, diffusion = S, state.variable=c("x1"),solve.variable=c("x1") )
    simulate_sde_on_branch(time,yuima.1d,X0,step)
}
\end{lstlisting}
and then the birth rate function (again the same as in Section \ref{secSimul})
\begin{lstlisting}
fbirth_rate_constrained<-function(x,params,...){
    x<-x[2]
    params$scale/(1+exp(-x))
}
\end{lstlisting}
It is important to have the \code{...} (i.e. the three dots following \code{params}
in the above code snippet) in the function's interface. This is because when called in the
package other parameters can be passed to it for generality, even though the user's implementation will not
require them. Having defined the functions we define the parameters under which we want to simulate
\begin{lstlisting}
true_sde.params<-list(a11=1,s11=1,psi1=0)
true_birth.params<-list(scale=5)
numtips<-200
tree_height<-max(15,log(numtips))
step<-0.001
\end{lstlisting}
If we had assumed a non--zero extinction rate, then its definition and parameters would be handled
in exactly the same way. With all of this we can simulate our trait--dependent speciation process
\begin{lstlisting}
simres<-simulate_phylproc(tree_height,simul.params=true_sde.params,X0=X0,fbirth=fbirth_rate_constrained,fdeath=NULL,fbirth.params=true_birth.params,fdeath.params=NULL,fsimulphenotype=simulate_OU_sde,n.contemporary=numtips,n.tips.total=100*numtips,step=step)
\end{lstlisting}
It is worth commenting about three values here, \code{tree\_height}, \code{n.contemporary} and \code{num.tips.total}.
The package first simulates the backbone lineage of length equal to \code{tree\_height}. Then, it will
start to simulate lineages coming out of the backbone lineage. The simulation needs a stopping condition, it will
either be that all birth events have taken place or \code{n.contemporary} tips (tips at height \code{tree\_height})
are generated, or \code{num.tips.total} are generated. The latter is for the situation when extinction is present. 
The value of \code{num.tips.total} will be the total number of tips, contemporary and extinct. Here it is just
given for illustrative purposes. The value of \code{step} is the simulation step size. 
After simulation if one wants to plot the trait trajectory over the tree one may use
\begin{lstlisting}
draw_phylproc(simres)
\end{lstlisting}
The figure is a bare drawing of the trait's evolution on the tree. Any plot components like axes have
to be added manually by the user. The tree can also be plotted, using e.g. \pkg{ape}'s plotting
capabilities
\begin{lstlisting}
plot(simres$tree,show.tip.label=FALSE,root.edge=TRUE)
\end{lstlisting}

In order to estimate parameters we need a phylogenetic tree and a matrix with tip measurements. The phylogeny
has to be in the \code{phylo} format. We can recover them from the simulated object as
\begin{lstlisting}
phyltree<-simres$tree
phenotypedata<-get_phylogenetic_sample(simres)
\end{lstlisting}
The package provides the functionality to recover the tip measurements using an inbuilt function 
\code{get\_phylogenetic\_sample()}.
We are now ready to perform the ABC inference, using some random starting parameters and choosing the 
distance calculation methods
\begin{lstlisting}
sde.params<-list(a11=5*runif(1),s11=5*runif(1),psi1=0,positivevars=c(TRUE,TRUE,FALSE),abcstepsd=rep(0.1,3))
birth.params<-list(scale=4+5*runif(1),maxval=10,abcstepsd=0.5,positivevars=c(TRUE,TRUE),fixed=c(FALSE,TRUE))
par0<-list(phenotype.model.params=sde.params,birth.params=birth.params)
X0<-c(0)
tree_dist<-"wRFnorm.dist"
data_dist<-"variancemean"
eps<-0.2
abcsteps<-1000
step<-0.001
ABCres<-PCM_ABC(phyltree=phyltree,phenotypedata=phenotypedata,par0=par0,phenotype.model=simulate_OU_sde,fbirth=fbirth,fdeath=NULL,X0=X0,step=step,abcsteps=abcsteps,eps=eps,tree.fixed=FALSE,dist_method=c(data_dist,tree_dist))}
\end{lstlisting}
If we wanted to assume a fixed tree (i.e. the phenotype does not affect speciation) we would set
\code{tree.fixed=FALSE} and e.g. \code{tree.dist<-NA}.

Afterwards (for the above setup it takes a bit under  one day in \proglang{R} $3.4.2$ for 
openSUSE $42.3$ (x$86$\_$64$) on a $3.50$GHz processor), we need to extract the estimated parameters.
The field \code{ABCres\$param.estimate}
is a list with two lists (one if  \code{tree.fixed=FALSE} or three if extinction present). The first list 
\code{phenotype.model.params} contains the point estimates of the trait evolution's process's parameters
and the second list \code{birth.params} the point estimates of the speciation rate's parameters.
If extinction is present a third list \code{death.params} with
the point estimates of the extinction rate's parameters will be present. The point
estimates are calculated as described in Alg. \ref{algABCpcmabc} as a weighted, by the inverse distances,
average of the accepted parameter values. The field \code{ABCres\$all.accepted} 
contains all the accepted parameters, including the distance from the observed data,
field \code{distance.from.data} for each accepted parameter set and also the 
inverse of the distance, field \code{inv.distance.from.data}. In the output object of 
\code{PCM\_ABC} are also two further fields: \code{sum.dists.from.data} the sum of the distances
from the observed data for all accepted parameters and \code{sum.inv.dists.from.data} the sum of the 
inverses of these distances.

\section{Conclusions}\label{secConc}
\subsection{The possibilities of the software}\label{sbsecPoss}
The \pkg{pcmabc} package is designed for maximum flexibility from the user's perspective.
It will be easiest to provide the full call of the function and discuss the more involved
components. This we did in the short tutorial in Section \ref{secTutorial}.

One could say that some of the parameters are extremely technical and maybe should
be hidden from the user. However, as a lot concerning the 
probabilistic/statistical properties of these models is not clear at this stage,
the user should have the possibility to experiment. Such experimentation should 
lead to better understanding of the underlying properties of the system. In turn,
this will allow us to know what is the best choice of these parameters (and to make 
them the default ones).

The package is extremely flexible and hence should be attractive for various types of studies.
Coupled with a model selection procedure it will allow for comparing different models 
of evolution and hence, asking questions about the system under study. 
The basic question one will want to ask is do the speciation dynamics
depend on the trait under study or not. For this one can try a constant birth--death
rate function, time dependent (but trait independent) speciation dynamics and 
trait dependent speciation dynamics. Treating time as a ``trait'' one may
study if the speciation dynamics increase, decrease with time or maybe
exhibit some sort of periodic behaviour. Similarly with trait dependent
dynamics---are the speciation rates monotonic w.r.t. the trait,
is there a carrying capacity, periodicity or maybe
some more complicated dynamics.

\subsection{Some limitations}\label{sbsecLimits}  
The one main feature that is lacking in the package at the moment is the possibility
to implement punctuated equilibrium models (e.g. \cite{KBar2014,FBok2002}), i.e. jumps at speciation points,
or more generally including direct feedback from the speciation dynamics into the trait dynamics.
This is as for computational efficiency we first simulate the whole lineage and only
afterwards, through the rejection sampling algorithm, simulate the points of speciation on this lineage. 
Hence, if the speciation event is meant to have an effect on the already simulated lineage
(e.g. through a jump) the already simulated trait data following the speciation point becomes invalid.
And this in turn could invalidate the thinning from the rejection sampling  algorithm and hence, the simulated
speciation event. Therefore, to include such models a different simulation algorithm has to be
developed.

Having written the above caveat our approach is still extremely general and it is important
to point out that it allows for another, biologically very relevant, type of speciation driven
evolution. Namely, the simulation algorithm has an inbuilt concept of a ``spine'' (or in other
words main) lineage. Speciation driven dynamics, like jumps at speciation points, can take place 
at the start of new lineages. This is consistent with the idea that a new lineage (species)
broke off because of some dramatic event, sudden jump in the trait.

Even though this is not directly evident from the user interface, the package
can easily handle time--heterogeneous models. What one passes to the simulation functions
is the trait value at the start of the branch or the trait value at the potential
branching event time. And then the simulation procedure evolves taking (from the package's
perspective) $0$ as the starting time of the branch. However, one can have time (from the root)
as one of the elements of the trait vector and then this can be used appropriately in the 
definition of the transition simulation procedures and birth--death rate functions. Therefore,
one can immediately recognize that one can include other dummy ``control'' traits, e.g.
environments, (geological) epochs, e.t.c. All that needs to be remembered is that
all user--defined functions have to appropriately treat these dimensions. Furthermore,
the package allows for defining fixed (i.e. not estimated) trait and speciation dynamics 
parameters, providing immense flexibility.

\subsection{Directions of development}\label{sbsecDev}
Despite the generality of the package there is a lot of space for further development
both in theoretical and implementation directions. A more involved and detailed study
is required to know what are the optimal inference settings, what distance measures 
should be used. For specific models one should ask what parameters are estimable.

From the perspective of the implementation, speciation dependent trait evolution is missing.
For effectiveness' sake full lineages are simulated and only then are speciation events
marked on them.  This implies that on the ``main'' lineage dependent
speciation dependent trait evolution cannot take place. Hence, new simulation
algorithms have to be developed that will not discard everything after 
the first time the branching influences the phenotype.

\section*{Acknowledgments}
We are grateful to an anonymous reviewer for their comments which have significantly improved the
manuscript.
KB's research is supported by the Swedish Research Council's (Vetenskapsr\aa det) grant no. $2017$--$04951$.

\bibliographystyle{plain}
\bibliography{pcmabc}

\end{document}